\documentclass[twocolumn]{aastex631}
\usepackage{graphicx,amsmath,amsthm,xspace,apjfonts}

\newcommand\lamser{$\lambda$\,Ser\xspace}
\newcommand{\numax}{\ensuremath{\nu_{\textrm{max}}}}
\newcommand{\dnu}{\ensuremath{\Delta\nu}}
\newcommand{\muHz}{\mbox{$\mu$Hz}}

\shorttitle{Exoplanet Host Star $\lambda$\,Serpentis}
\shortauthors{Metcalfe et al.}

\journalinfo{The Astronomical Journal}

\begin{document}

\title{\large Asteroseismology and Spectropolarimetry of the Exoplanet Host Star $\lambda$~Serpentis}

\suppressAffiliations

\author[0000-0003-4034-0416]{Travis S.~Metcalfe}
\affiliation{White Dwarf Research Corporation, 9020 Brumm Trail, Golden, CO 80403, USA}

\author[0000-0002-1988-143X]{Derek Buzasi} 
\affiliation{Department of Chemistry and Physics, Florida Gulf Coast University, 10501 FGCU Blvd S, Fort Myers, FL 33965, USA}

\author[0000-0001-8832-4488]{Daniel Huber} 
\affiliation{Institute for Astronomy, University of Hawai`i, 2680 Woodlawn Drive, Honolulu, HI 96822, USA}
\affiliation{Sydney Institute for Astronomy (SIfA), School of Physics, University of Sydney, Camperdown, NSW 2006, Australia}

\author[0000-0002-7549-7766]{Marc H.~Pinsonneault} 
\affiliation{Department of Astronomy, The Ohio State University, 140 West 18th Avenue, Columbus, OH 43210, USA}

\author[0000-0002-4284-8638]{Jennifer L.~van~Saders} 
\affiliation{Institute for Astronomy, University of Hawai`i, 2680 Woodlawn Drive, Honolulu, HI 96822, USA}

\author[0000-0002-1242-5124]{Thomas R.~Ayres} 
\affiliation{Center for Astrophysics and Space Astronomy, 389 UCB, University of Colorado, Boulder, CO 80309, USA}

\author[0000-0002-6163-3472]{Sarbani Basu} 
\affiliation{Department of Astronomy, Yale University, PO Box 208101, New Haven, CT 06520-8101, USA}

\author[0000-0002-0210-2276]{Jeremy J.~Drake} 
\affiliation{Harvard-Smithsonian Center for Astrophysics, Cambridge, MA 02138, USA}

\author[0000-0002-4996-0753]{Ricky Egeland} 
\affiliation{White Dwarf Research Corporation, 9020 Brumm Trail, Golden, CO 80403, USA}

\author[0000-0003-3061-4591]{Oleg Kochukhov} 
\affiliation{Department of Physics and Astronomy, Uppsala University, Box 516, SE-75120 Uppsala, Sweden}

\author[0000-0001-7624-9222]{Pascal Petit}
\affiliation{Universit\'e de Toulouse, CNRS, CNES, 14 avenue Edouard Belin, 31400, Toulouse, France}

\author[0000-0001-7032-8480]{Steven H.~Saar} 
\affiliation{Harvard-Smithsonian Center for Astrophysics, Cambridge, MA 02138, USA}

\author[0000-0001-5986-3423]{Victor~See} 
\altaffiliation{ESA Research Fellow}
\affiliation{European Space Agency (ESA), European Space Research and Technology Centre (ESTEC), Keplerlaan 1, 2201 AZ Noordwijk, The Netherlands}

\author[0000-0002-3481-9052]{Keivan G.~Stassun} 
\affiliation{Vanderbilt University, Department of Physics \& Astronomy, 6301 Stevenson Center Lane, Nashville, TN 37235, USA}

\author[0000-0003-3020-4437]{Yaguang Li} 
\affiliation{Sydney Institute for Astronomy (SIfA), School of Physics, University of Sydney, Camperdown, NSW 2006, Australia}

\author[0000-0001-5222-4661]{Timothy R.~Bedding} 
\affiliation{Sydney Institute for Astronomy (SIfA), School of Physics, University of Sydney, Camperdown, NSW 2006, Australia}

\author[0000-0003-0377-0740]{Sylvain N.~Breton} 
\affiliation{INAF – Osservatorio Astrofisico di Catania, Via S. Sofia, 78, 95123 Catania, Italy}

\author[0000-0002-3020-9409]{Adam J.~Finley} 
\affiliation{Universit\'e Paris-Saclay, Universit\'e Paris Cit\'e, CEA, CNRS, AIM, 91191, Gif-sur-Yvette, France}

\author[0000-0002-8854-3776]{Rafael A.~Garc\'\i a} 
\affiliation{Universit\'e Paris-Saclay, Universit\'e Paris Cit\'e, CEA, CNRS, AIM, 91191, Gif-sur-Yvette, France}

\author[0000-0002-9037-0018]{Hans Kjeldsen} 
\affiliation{Stellar Astrophysics Centre, Aarhus University, Ny Munkegade 120, DK-8000 Aarhus C, Denmark}

\author[0000-0001-9169-2599]{Martin B.~Nielsen} 
\affiliation{School of Physics \& Astronomy, University of Birmingham, Edgbaston, Birmingham B15 2TT, UK}

\author[0000-0001-7664-648X]{J.~M.~Joel Ong} 
\altaffiliation{NASA Hubble Fellow}
\affiliation{Institute for Astronomy, University of Hawai`i, 2680 Woodlawn Drive, Honolulu, HI 96822, USA}

\author[0000-0001-9234-430X]{Jakob L.~R{\o}rsted} 
\affiliation{Stellar Astrophysics Centre, Aarhus University, Ny Munkegade 120, DK-8000 Aarhus C, Denmark}

\author[0000-0002-5496-365X]{Amalie Stokholm} 
\affiliation{Dipartimento di Fisica e Astronomia, Universit\`{a} degli Studi di Bologna, Via Gobetti 93/2, I-40129 Bologna, Italy}
\affiliation{INAF -- Osservatorio di Astrofisica e Scienza dello Spazio di Bologna, Via Gobetti 93/3, I-40129 Bologna, Italy}
\affiliation{Stellar Astrophysics Centre, Aarhus University, Ny Munkegade 120, DK-8000 Aarhus C, Denmark}

\author[0000-0003-1687-3271]{Mark L.~Winther} 
\affiliation{Stellar Astrophysics Centre, Aarhus University, Ny Munkegade 120, DK-8000 Aarhus C, Denmark}

\author[0000-0002-2361-5812]{Catherine A.~Clark} 
\affiliation{Jet Propulsion Laboratory, California Institute of Technology, Pasadena, CA 91109 USA}
\affiliation{NASA Exoplanet Science Institute, IPAC, California Institute of Technology, Pasadena, CA 91125 USA}

\author[0000-0003-4556-1277]{Diego Godoy-Rivera} 
\affiliation{Instituto de Astrof\'{\i}sica de Canarias, E-38205 La Laguna, Tenerife, Spain}
\affiliation{Universidad de La Laguna, Departamento de Astrofísica, E-38206 La Laguna, Tenerife, Spain}

\author[0000-0002-0551-046X]{Ilya V.~Ilyin} 
\affiliation{Leibniz-Institut f\"ur Astrophysik Potsdam (AIP), An der Sternwarte 16, D-14482 Potsdam, Germany}

\author[0000-0002-6192-6494]{Klaus G.~Strassmeier} 
\affiliation{Leibniz-Institut f\"ur Astrophysik Potsdam (AIP), An der Sternwarte 16, D-14482 Potsdam, Germany}

\author[0000-0003-2490-4779]{Sandra V.~Jeffers} 
\affiliation{Max-Planck-Institut f\"ur Sonnensystemforschung, Justus-von-Liebig-weg 3, 37077, G\"ottingen, Germany}

\author[0000-0001-5522-8887]{Stephen C.~Marsden} 
\affiliation{Centre for Astrophysics, University of Southern Queensland, Toowoomba, Queensland, 4350, Australia}

\author[0000-0001-5371-2675]{Aline A.~Vidotto} 
\affiliation{Leiden Observatory, Leiden University, PO Box 9513, 2300 RA, Leiden, The Netherlands}

\author{Sallie Baliunas} 
\affiliation{Harvard-Smithsonian Center for Astrophysics, Cambridge, MA 02138, USA}

\author{Willie Soon} 
\affiliation{Institute of Earth Physics and Space Science (EPSS), Sopron, Hungary}
\affiliation{Center for Environmental Research and Earth Sciences, Salem, MA 01970, USA}

\begin{abstract}

The bright star \lamser hosts a hot Neptune with a minimum mass of 13.6~$M_\oplus$ and a 
15.5 day orbit. It also appears to be a solar analog, with a mean rotation period of 25.8 
days and surface differential rotation very similar to the Sun. We aim to characterize 
the fundamental properties of this system, and to constrain the evolutionary pathway that 
led to its present configuration. We detect solar-like oscillations in time series 
photometry from the Transiting Exoplanet Survey Satellite (TESS), and we derive precise 
asteroseismic properties from detailed modeling. We obtain new spectropolarimetric data, 
and we use them to reconstruct the large-scale magnetic field morphology. We reanalyze 
the complete time series of chromospheric activity measurements from the Mount Wilson 
Observatory, and we present new X-ray and ultraviolet observations from the Chandra and 
Hubble space telescopes. Finally, we use the updated observational constraints to assess 
the rotational history of the star and to estimate the wind braking torque. We conclude 
that the remaining uncertainty on stellar age currently prevents an unambiguous 
interpretation of the properties of \lamser, and that the rate of angular momentum loss 
appears to be higher than for other stars with similar Rossby number. Future 
asteroseismic observations may help to improve the precision of the stellar age.

\end{abstract}

\keywords{Spectropolarimetry; Stellar activity; Stellar evolution; Stellar oscillations; Stellar rotation}

\NewPageAfterKeywords
\section{Introduction}\label{sec1}

Asteroseismology and spectropolarimetry are powerful tools to study magnetic stellar 
evolution. The surface convective regions of Sun-like stars generate sound waves over a 
broad range of frequencies, some of which are resonant inside the spherical cavity of the 
star and set up standing waves that produce tiny variations in brightness. These natural 
oscillations probe the interior conditions of the star, and can be used to infer basic 
physical properties including the stellar radius, mass, and age 
\citep[see][]{GarciaBallot2019}. Surface magnetism can break the spherical symmetry of 
the stellar atmosphere, polarizing the radiated starlight in a way that encodes 
information about the strength and orientation of the global magnetic field. Multiple 
snapshot observations of the disk-integrated polarization signature as a star rotates can 
be used to reconstruct the complete morphology of the large-scale field 
\citep[see][]{Kochukhov2016}, which sculpts the escaping stellar wind and influences the 
rate of angular momentum loss. When combined, these two methods can provide important new 
constraints on how the magnetic properties of solar-type stars change throughout their 
lifetimes.

Very few previous studies have combined information from asteroseismic and 
spectropolarimetric observations. Both techniques require extremely precise measurements, 
which have only recently become available from space-based photometry \citep{Borucki2010, 
Ricker2014}, and ground-based spectropolarimetry \citep{Marsden2014, Strassmeier2015}. 
Early efforts relied on observations of massive stars with fossil magnetic fields 
\citep{MathisNeiner2015}, while more recent work has concentrated on the magnetic 
evolution of solar-type stars \citep{Metcalfe2021, Metcalfe2022, Metcalfe2023}. The 
latter studies have begun to probe the physical mechanisms that may be responsible for 
the onset of weakened magnetic braking \citep{vanSaders2016, Hall2021}, revealing a 
dramatic decrease in the wind braking torque during the second half of main sequence 
lifetimes. Here we use these techniques to investigate an old main sequence star with 
some unusual properties.

The exoplanet host star \lamser has been studied for decades as an old solar analog. 
Long-term observations of its chromospheric emission suggest a nearly constant activity 
level comparable to recent solar minima \citep{Baliunas1995}, while higher cadence 
measurements reveal a mean rotation period and surface differential rotation that are 
both similar to the Sun \citep{Donahue1996}. It has an unusually high lithium abundance 
for an old solar analog \citep[A(Li)=1.96;][]{XingXing2012}, with an enhancement 
comparable to HD\,96423 which might be explained by planetary engulfment 
\citep{Carlos2016}. It was recently confirmed to host a hot Neptune with a minimum mass 
of 13.6~$M_\oplus$ in a 15.5 day orbit \citep{Rosenthal2021}. In this paper, we aim to 
characterize the properties of \lamser and to constrain the evolutionary pathway that led 
to its present configuration. In Section~\ref{sec2} we describe new asteroseismic and 
spectropolarimetric observations, new measurements in the X-ray and ultraviolet, as well 
as a reanalysis of archival chromospheric activity data. In Section~\ref{sec3} we derive 
precise stellar properties from asteroseismic modeling, we infer the global magnetic 
morphology from Zeeman Doppler Imaging (ZDI), and we attempt to interpret these 
measurements in the context of rotational and magnetic evolution. Finally in 
Section~\ref{sec4}, we discuss possible scenarios to explain the unusual properties of 
\lamser and we outline future measurements that might clarify its evolutionary status.

\section{Observations}\label{sec2}

\subsection{TESS Photometry}\label{sec2.1} 

The Transiting Exoplanet Survey Satellite (TESS) observed \lamser during Sector 51 (2022 
April 22 -- 2022 May 18) in 20~s cadence, which has been demonstrated to have superior 
photometric precision to 2~min data for bright stars \citep{huber22}. We were able to 
improve on the standard Science Processing Operations Center (SPOC) data product by using 
our own method to extract the light curve \citep{Nielsen2020}. We approached the postage 
stamp image pixel-by-pixel, extracting a time series for each pixel. We then took the 
brightest pixel, which was also the closest to the nominal position of the star, as our 
initial time series. The pixel time series quality figure of merit was parameterized by 
\begin{equation} 
{q = \sum_{i=1}^{N-1}\mid f_{i+1}-f_{i}\mid}, 
\end{equation} 
where $f_i$ is the flux at cadence $i$, and $N$ is the length of the time series. Using 
the first differences of the light curve acts to whiten the time series, and thus correct 
for its non-stationary nature \citep{Nason2006}; similar approaches have been used in 
astronomical time series analysis by \cite{garcia11}, \cite{Buzasi2015}, \cite{Prsa2019}, 
and \cite{Nielsen2020}, among other authors. We then added the light curve from the pixel 
which most decreased our figure of merit, and repeated the process until the light curve 
quality as measured by our quality measure $q$ ceased to improve. The resulting pixel 
collection was adopted as our aperture mask. Finally we detrended the light curve 
produced using this mask against the centroid pixel coordinates by fitting a second-order 
polynomial with cross terms. Similar approaches have been used for K2 data reduction 
\citep[e.g., see][]{Vanderburg2014}.

\begin{figure*}
\centering\includegraphics[width=\textwidth,trim=0 10 0 0,clip]{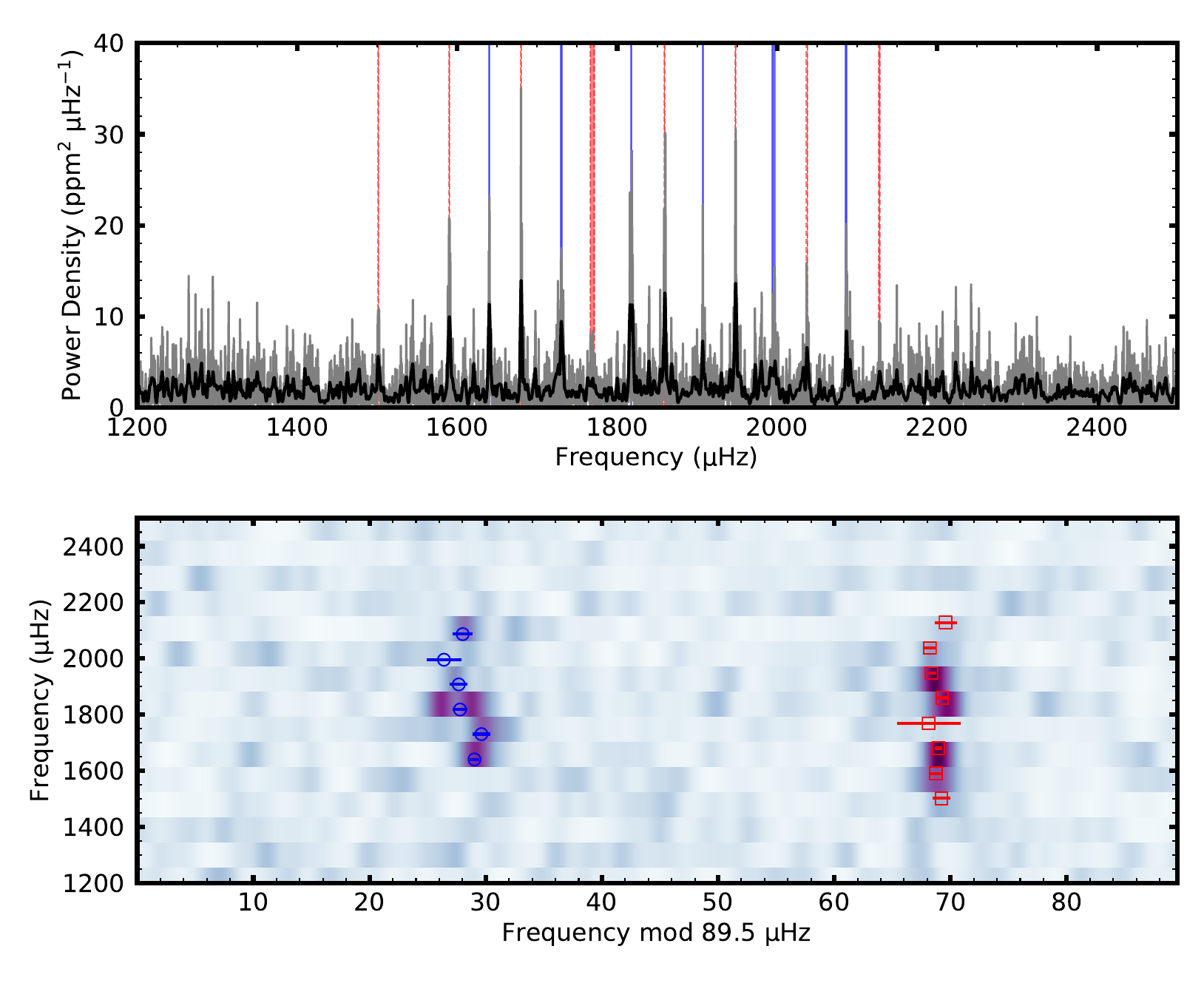}
\caption{Power spectrum (top) and \'{e}chelle diagram (bottom) centered on the power 
excess due to solar-like oscillations detected in \lamser. Blue solid lines and circles 
indicate extracted radial ($l=0$) modes, while red dashed lines and squares show 
extracted dipole ($l=1$) modes.\vspace*{12pt}\label{fig1}}
\end{figure*}

The overall noise level of our light curve, measured point-to-point, is approximately 7\% 
better than that of the SPOC product, and we were able to recover more points: 67,491 as 
compared to 55,445 in the SPOC light curve (or 58,528 if requirements are relaxed to 
include points with nonzero quality flags). This improved the duty cycle from 52-55\% to 
almost 64\%. The top panel of Figure~\ref{fig1} shows the power spectrum of the resulting 
time series with clear evidence for solar-like oscillations centered near 1900~\muHz. We 
use this same approach in Appendix~\ref{appA} to quantify non-detections of oscillations 
in eight additional TESS targets.

To extract individual frequencies, four different groups of coauthors applied either 
iterative sine-wave fitting \citep[e.g.][]{lenz05,kjeldsen05,bedding07} or Lorentzian 
mode-profile fitting \citep[e.g.][]{garcia09, handberg11, appourchaux12, mosser11c, 
corsaro14, corsaro15, liyg20, breton22}. For each mode, we required at least two 
independent methods to return the same frequency within uncertainties. For the final list 
we adopted values from a single method, with uncertainties derived by adding in 
quadrature the median formal uncertainty and the standard deviation of the extracted 
frequencies from all methods that identified a given mode.

The bottom panel of Figure~\ref{fig1} shows an \'{e}chelle diagram with a large 
separation of $\dnu=89.5$~\muHz\ and the extracted frequencies. We identify six radial 
($l=0$) and eight dipole ($l=1$) modes, but we were unable to identify any quadrupole 
($l=2$) modes with confidence. Mode identification was confirmed using well-known 
patterns between $\dnu$ and the frequency offset $\epsilon$ \citep{white11} and by 
comparison with similar stars (e.g., KIC~7296438) from the Kepler LEGACY sample 
\citep{lund17b}. The final frequency list is shown in Table~\ref{tab1}, providing a 
primary input for the asteroseismic modeling described in Section~\ref{sec3.1}.\vfill

\subsection{Spectropolarimetry}\label{sec2.2} 

\begin{deluxetable}{ccc}[t]
\setlength{\tabcolsep}{28pt}
\tablecaption{Identified oscillation frequencies for \lamser.\label{tab1}}
\tablehead{\colhead{$l$} & \colhead{$\nu~(\muHz)$} & \colhead{$\sigma_{\nu}~(\muHz)$}}
\startdata
0 & 1640.04 & 0.42 \\
0 & 1730.13 & 0.74 \\
0 & 1817.79 & 0.61 \\
0 & 1907.18 & 0.77 \\ 
0 & 1995.41 & 1.48 \\
0 & 2086.52 & 0.85 \\ 
1 & 1501.22 & 0.74 \\
1 & 1590.24 & 0.50 \\
1 & 1679.97 & 0.32 \\
1 & 1768.62 & 2.72 \\
1 & 1859.28 & 0.60 \\
1 & 1947.86 & 0.42 \\
1 & 2037.22 & 0.52 \\
1 & 2128.08 & 0.94 \\
\enddata
\end{deluxetable}
\vspace*{-12pt}

\begin{figure}
\centering\includegraphics[width=\linewidth]{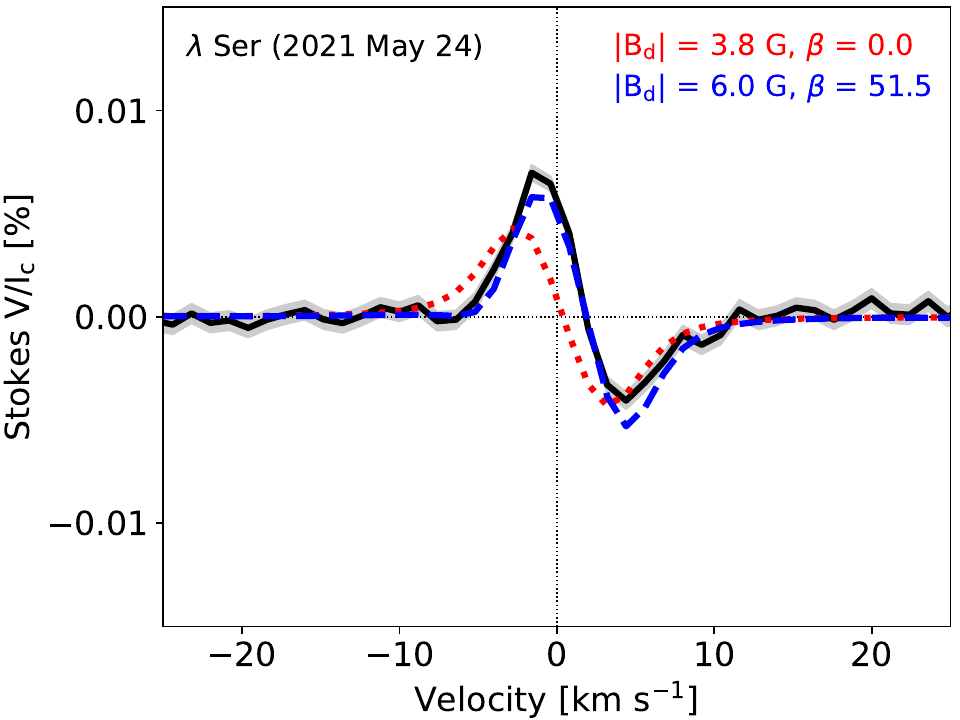}
\caption{Stokes~$V$ polarization profile for \lamser from LBT observations on 
2021~May~24. The mean profile is shown as a black line, with uncertainties indicated by 
the gray shaded area. The red and blue lines are model profiles assuming dipole
geometry and fixed inclination $i=50^\circ$ with different obliquity angles $\beta$.
\label{fig2}}
\end{figure}

Spectropolarimetric observations of \lamser were obtained on 2021 May 24 using the 
Potsdam Echelle Polarimetric and Spectroscopic Instrument 
\citep[PEPSI;][]{Strassmeier2015} installed at the $2\times8.4$-m Large Binocular 
Telescope (LBT). The instrumental setup, resulting in $R$\,=\,130,000 observations over 
the 475--540~nm and 623--743~nm wavelength regions, and the data reduction procedures 
were the same as described in \citet{Metcalfe2019b}. Considering the low-amplitude 
polarization signal, a multi-line method is necessary to achieve a magnetic field 
detection. Here we employed the least-squares deconvolution \citep[LSD;][]{Kochukhov2010} 
technique to derive high-quality intensity and circular polarization profiles. The line 
mask necessary for this analysis was constructed from the output of an ``extract 
stellar'' request to the VALD database \citep{Ryabchikova2015} with stellar atmospheric 
parameters from \citet{Brewer2016}. This line list contained 1300 metal lines deeper than 
10\% of the continuum in the wavelength region covered by our PEPSI data. Combining 
information from these lines yielded an LSD Stokes $V$ profile with an uncertainty of 
4.6~ppm, which showed a clear magnetic signal (see Figure~\ref{fig2}). We measured a mean 
longitudinal magnetic field $\left<B_{\rm z}\right>=0.674\pm0.048$~G from these 
observations and estimated the strength of an axisymmetric dipole magnetic field to be 
$B_{\rm d}=3.8$~G using the line profile modeling technique described in 
\citet{Metcalfe2019b} and adopting $i=50\degr$ (see Section~\ref{sec3.2}). However, 
the resulting synthetic profile (dotted red line in Figure~\ref{fig2}) did not 
provide an adequate description of the observations, suggesting the presence of 
non-axisymmetric global field components. Consequently, we generalized the modeling by 
allowing an inclined dipole geometry. This produced a better fit to the observed 
Stokes~$V$ profile with $B_{\rm d}=6.0$~G and a magnetic obliquity of $\beta=51.5\degr$ 
(dashed blue line in Figure~\ref{fig2}).

Additional spectropolarimetric observations of \lamser covering a broad range of rotation 
phases were collected with Neo-NARVAL during the summer of 2021, allowing us to model the 
detailed morphology of its large-scale magnetic field. The complete data set included 19 
visits to the star, secured between 2021 July 7 and 2021 August 17, with a maximum of one 
observation per night. Neo-NARVAL is an upgrade to the NARVAL instrument at T\'elescope 
Bernard Lyot \citep[TBL;][]{Auriere2003}. Neo-NARVAL echelle spectra collect a broad 
optical wavelength region (370--1,000~nm) in a single frame, with a spectral resolution 
close to $R$\,=\,65,000. Every polarimetric sequence is obtained from the combination of 
four exposures taken with the two half-wave Fresnel rhombs rotated about the optical axis 
\citep{Semel1993}. Each polarized sequence provides simultaneous access to a Stokes~$I$ 
spectrum and another Stokes parameter (circular or linear polarization). The data set 
gathered for \lamser was restricted to the Stokes~$V$ parameter, since the amplitude of 
Zeeman signatures is expected to be largest in circular polarization 
\citep{deglinnocenti1992}. Each sequence also provides a ``null'' spectrum, which should 
contain only noise and serves as a diagnostic of possible instrumental or stellar 
contamination to the polarized spectrum.

The Neo-NARVAL upgrade, installed in 2019, consisted of a new detector and enhanced 
velocimetric capabilities \citep{LopezAriste2022}. At the time of our observations, the 
instrument suffered from a loss of flux in the bluest orders, later identified as a fiber 
link issue. The new reduction pipeline \citep{LopezAriste2022} provided an unsatisfactory 
extraction of spectral orders affected by a very low signal-to-noise ratio (S/N), so we 
discarded from our reduced data all spectral bins with a wavelength below 470~nm. The LSD 
analysis and interpretation of these data are described in Section~\ref{sec3.2}.

\subsection{X-Ray Measurements}\label{sec2.3}

We observed \lamser with the Chandra High Resolution Camera-Imaging detector (HRC-I; 
ObsID 22307) on 2020 April 25 in a single pointing with a net exposure time of 6,103\,s. 
This instrument was preferred over the Advanced CCD Imaging Spectrometer (ACIS) because a 
growing contamination layer on the ACIS optical blocking filter severely curtails the 
low-energy response below about 1~keV. The HRC-I data were reprocessed using the Chandra 
Interactive Analysis of Observations \citep[CIAO;][]{Fruscione2006} software 
version 4.15 and calibration database version 4.10.2. Since the HRC-I has essentially no 
intrinsic energy resolution, the analysis entailed examining the source photon event list 
for significant variability, extracting photon events attributed to \lamser, and 
converting the observed count rate into a source flux.

\begin{figure}
\centering\includegraphics[width=\linewidth,trim=0 0 0 20,clip]{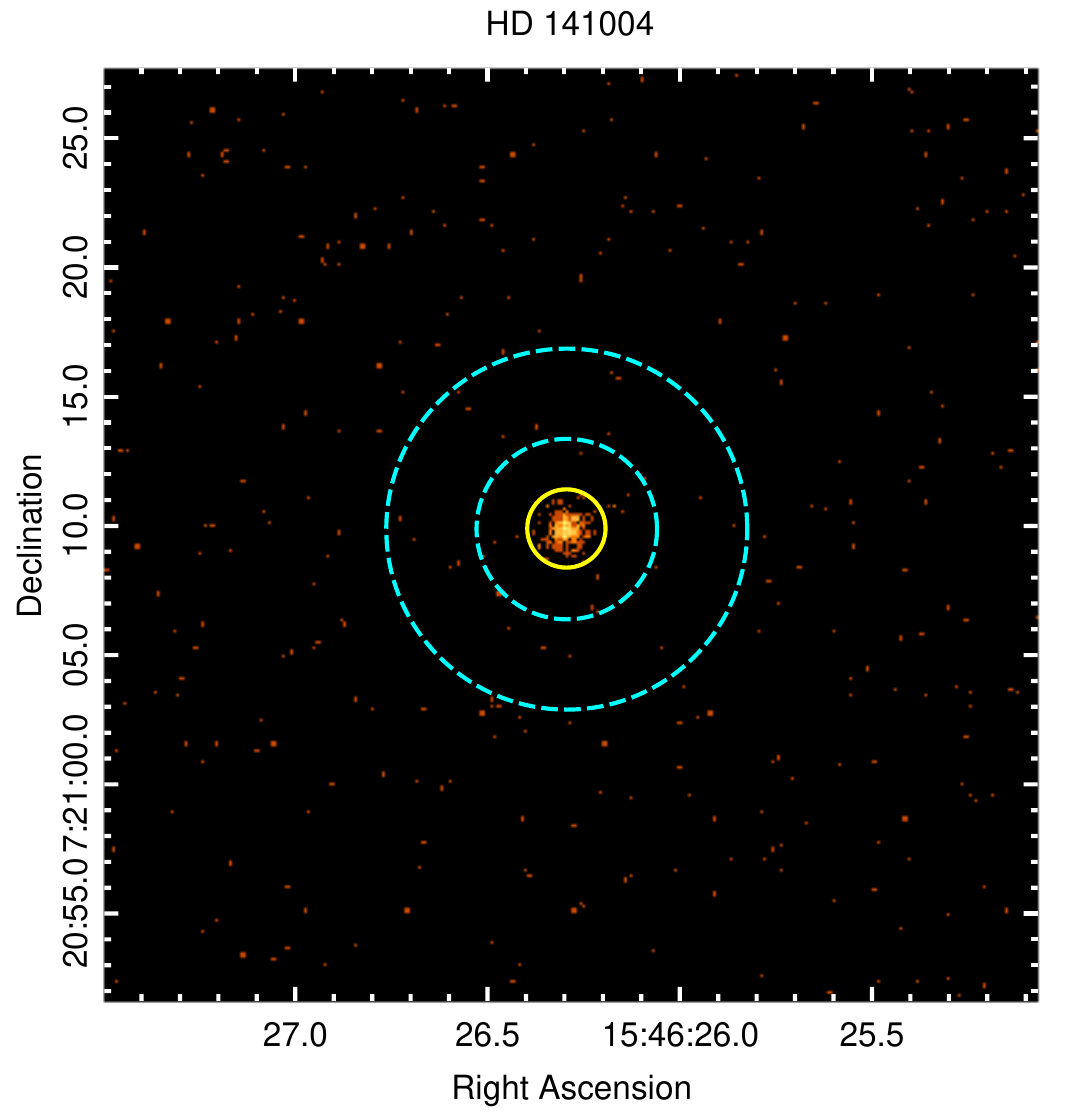}
\caption{Chandra HRC-I image of \lamser illustrating the source and innermost of two 
background regions used for the count rate measurement (see Section~\ref{sec2.3} for 
details).\label{fig3}}
\end{figure}

An image of the detected events in the vicinity of \lamser is shown in Figure~\ref{fig3}. 
Overlaid is the adopted source extraction region (yellow), together with the innermost of 
two background regions (cyan) employed to estimate the background signal. The source 
region was placed at the centroid of the detected events and had a radius of 1.5\arcsec, 
which corresponds to an encircled energy fraction of 95\%. The background region 
illustrated in Figure~\ref{fig3} was an annulus with inner and outer radii of 3.5\arcsec 
and 7\arcsec, respectively, centered on the source. We also estimated the background rate 
using a much larger annulus, covering the radius interval 130\arcsec--165\arcsec, to 
check for the presence of spatial variations in the background. In both cases the net 
source count rate was $0.049\pm 0.003$~count~s$^{-1}$ after correction for the encircled 
energy fraction. The extracted source counts were examined for variability using the 
Gregory-Loredo algorithm \citep{GregoryLoredo1992}; no significant variability was 
detected.

In addition to the Chandra observation, \lamser was observed in soft X-rays during the 
ROSAT era, initially as part of the all-sky survey (scans were acquired in 1990 August), 
and later during the pointed phase of the Guest Investigator program (a PSPCB exposure in 
1997 February, toward the end of the mission). Count rates (CR) for the two ROSAT 
observations were taken from facility catalogs hosted by the High-Energy Science and 
Archive Research Center (HEASARC), at the NASA Goddard Space Flight Center, as accessed 
through W3browse\footnote{\url{https://heasarc.gsfc.nasa.gov/cgi-bin/W3Browse/w3browse.pl}}. 
The {\tt rass2rxs} catalog listed CR\,=\,$0.083 \pm 0.016$~count~s$^{-1}$ for a sky 
survey exposure of 511~s. The pointings catalog {\tt rospspctotal} reported CR\,=\,$0.072 
\pm 0.006$~count~s$^{-1}$ for an exposure of 2.73~ks. The energy bandpass is the ROSAT 
standard, 0.1--2.4~keV. Documentation for these databases can be obtained through 
W3browse.

\begin{figure*}
\centering\includegraphics[width=\textwidth,trim=10 10 10 25,clip]{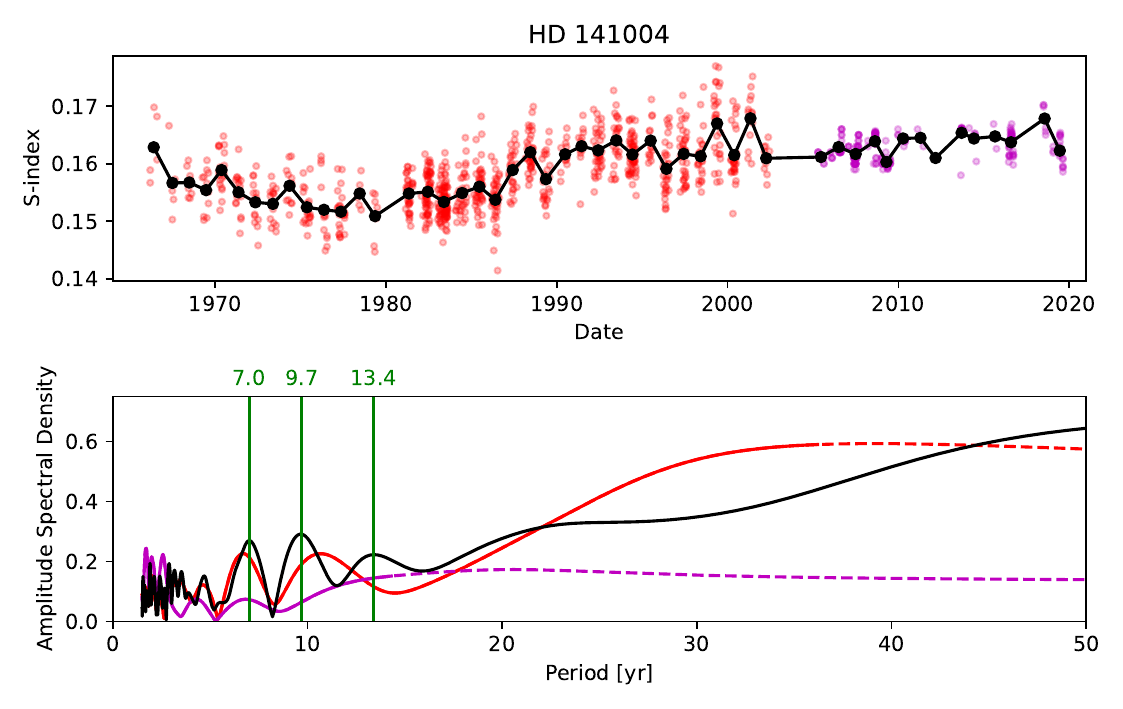}
\caption{\emph{Top:} 54-year $S$-index time series for \lamser. Observations are shown 
from MWO HKP (red) and Keck HIRES-2 (purple) with seasonal means (black). \emph{Bottom:} 
Lomb-Scargle periodogram of the time series, with MWO-only in red, Keck-only in purple, 
and the composite time series in black.  Dashed lines indicate portions of the 
periodogram beyond the duration of the time series.  The top three peaks in the composite 
time series are indicated with green vertical lines.  The periodogram is expressed in 
units of signal amplitude normalized by a standard deviation.\label{fig4}}
\end{figure*}

The count rates from ROSAT and Chandra were converted to X-ray fluxes at Earth using a 
method to derive an optimum energy conversion factor (ECF) for each camera system. The 
approach applied a sequence of coronal emission-measure distributions (EMD), convolved 
with detector-dependent sensitivity curves, to calculate X-ray surface fluxes for the 
target; finding the optimum EMD realization that achieved consistency between the 
calculated and model surface fluxes (each EMD level corresponds to a specific predicted 
X-ray surface flux). The approach is described by \cite{Ayres2022}, and is based on the 
empirical EMD models derived by \cite{Wood+2018} from Chandra Low-Energy Transmission 
Grating spectra of nearly two dozen F--M dwarfs. Detector sensitivity curves were 
calculated for the reference $\log{T}$ grid of each EMD using 
WebPIMMS\footnote{\url{https://cxc.harvard.edu/toolkit/pimms.jsp}}, for the unabsorbed 
0.1--2.4~keV X-ray flux, a solar-abundance APEC plasma model, and an interstellar column 
of $1{\times}10^{18}$~cm$^{-2}$, appropriate for a nearby ($d=11.9$~pc) star. There was 
good consistency among the three independent X-ray fluxes: $f_{\rm X}= 4.0 \pm 0.4 
{\times}10^{-13}$~erg~cm$^{-2}$~s$^{-1}$. At the Gaia distance, this corresponds to 
$L_{\rm X}=6.8 \pm 0.6 {\times}10^{27}$~erg~s$^{-1}$, or $\log{L_{\rm X}}$\,=\,27.83, 
about seven times larger than the sunspot-cycle-average Sun \citep{Ayres2022}.

We can estimate the mass-loss rate of \lamser by combining the X-ray luminosity 
determined above with the stellar radius inferred from asteroseismology (see 
Section~\ref{sec3.1}). For stars with mass-loss rates determined directly from Ly$\alpha$ 
measurements and other techniques, there is an empirical relation between the mass-loss 
rate and the X-ray flux per unit surface area, $\dot M \propto F_{\rm X}^{0.77}$ 
\citep{Wood2021}. The resulting estimate is slightly above the solar value, $\dot M = 1.6 
\pm 0.2\ \dot M_\odot$.

\subsection{Chromospheric Activity Data}\label{sec2.4}

We used synoptic observations of the $S$-index of chromospheric activity from the Mount 
Wilson Observatory (MWO) HK Project \citep{Wilson1978, Baliunas1996} and the Keck HIRES 
spectrometer \citep{Baum2022} to measure the rotation period of \lamser and to 
characterize its long-term magnetic variability. The $S$-index was defined by the MWO HK 
Photometer (HKP) and measured the ratio of emission from 0.1\,nm cores of the 
chromospheric Ca {\sc ii} H \& K lines to the sum of two nearby 2\,nm pseudo-continuum 
bandpasses \citep{Vaughan1978}.  This long-used proxy for stellar magnetic activity 
reveals the presence of decadal-scale cycles in the Sun \citep[e.g.][]{White1981, 
Egeland2017b} and Sun-like stars \citep[e.g.][]{Baliunas1995, Hall2007, Egeland2017}.  
The passage of surface active regions modulates the $S$-index such that when sampled at a 
sufficient cadence the stellar rotation period can be obtained \citep{Baliunas1983, 
Baliunas1996, Donahue1996}. The MWO HK Project observed \lamser from its inception in 
1966 until its termination in 2003. Previous studies have reported results on partial 
records of \lamser, but here we analyze the complete time series obtained by MWO.

\begin{deluxetable*}{lrrrrrrrr}
\tablecaption{Seasonal Rotation Period Detections for \lamser \label{tab2}}
\tablehead{\colhead{} & \multicolumn{4}{c}{Donahue} & \multicolumn{4}{c}{This Work} \\
\colhead{Season} & \colhead{$N_{\rm obs}$} & \colhead{$P_{\rm rot}$} & \colhead{$\Delta P$} & \colhead{FAP} & \colhead{$N_{\rm JD}$} & \colhead{$P_{\rm rot}$} & \colhead{$\sigma_P$} & \colhead{FAP}}
\startdata
  1970.36 &     &      &     &          & 21 & 19.4 & 0.7 & 3.01\%  \\
  1977.36 &     &      &     &          & 24 & 23.0 & 0.5 & 0.56\%  \\
  1981.34 & 103 & 24.4 & 0.2 & 0.0015\% & 35 & 25.5 & 0.5 & 0.003\% \\
  1983.38 & 271 & 27.1 & 0.3 & 3.4\%    &    &      &     &         \\
  1986.40 & 124 & 28.8 & 0.4 & 0.20\%   & 42 & 28.0 & 1.3 & 2.97\%  \\
  1987.41 &  84 & 24.3 & 0.3 & 2.1\%    &    &      &     &         \\
  1988.45 & 113 & 26.4 & 0.4 & 0.12\%   & 37 & 26.3 & 0.6 & 2.59\%  \\
  1992.42 &  91 & 23.6 & 0.4 & 1.5\%    & 39 & 23.3 & 0.2 & 0.04\%  \\\hline
  {\bf N detections}       & & 6    & & & & 6    & & \\
  {\bf Min $P_{\rm rot}$}  & & 23.6 & & & & 19.4 & & \\
  {\bf Max $P_{\rm rot}$}  & & 28.8 & & & & 28.0 & & \\
  {\bf Mean $P_{\rm rot}$} & & 25.8 & & & & 24.3 & & \\
\enddata
\tablecomments{\citeauthor{Donahue1993} values are taken from Table B.25 of \citet{Donahue1993}.  The decimal year in the Season column gives the mean decimal year of the observations in that season.  $N_{\rm obs}$ refers to the individual MWO observations analyzed by \citeauthor{Donahue1993}, while $N_{\rm JD}$ refers to the nightly (Julian Day) averages used in this work.}
\vspace*{-12pt}
\end{deluxetable*}
\vspace*{-24pt}

We extend the MWO observations using Keck Observatory HIRES-2 data obtained for the 
California Planet Search (CPS) and published in \citet{Baum2022}. These High Resolution 
Echelle Spectrometer (HIRES) observations (R $\sim$ 67,000) cover the Ca~{\sc ii} H \& K 
region, and the $S$-index is obtained by reducing the spectra and integrating the 
bandpasses of the MWO HKP-2 spectrophotometer \citep{Vaughan1978, Isaacson2010}.  While 
some data from \citet{Baum2022} were adjusted with a constant shift to be consistent with 
MWO, no such shift was applied to the data set for \lamser. The composite data set is 
shown in the top panel of Figure~\ref{fig4} with the seasonal means indicated. Some 
low-frequency variation is apparent by visual inspection of the seasonal means.  
Activity falls from the beginning of the observations in 1966 ($S = 0.1629$) until the 
global minimum seasonal mean in 1979 ($S = 0.1509$). From there, activity rises until 
about 1988 ($S = 0.1620$) and remains relatively constant thereafter. The global maximum 
seasonal mean occurs in 2001 ($S = 0.1679$), but this appears to be an intermittent 
outburst reaching levels comparable to 1999 and 2018. The standard deviation of the 
seasonal means including and after the 1988 season is $\sigma_S = 0.00236$ ($N = 29$), 
which is about half of the standard deviation for the whole series ($\sigma_S = 0.00460, 
N = 50$).  For reference, solar minimum has $S = 0.1621$ and the mean cycle maximum is $S 
= 0.177$ \citep{Egeland2017b}.

We employed the Lomb-Scargle periodogram \citep{Lomb1976,Scargle1982,Horne1986} to search 
for periodicity in the separate (MWO, Keck) and composite time series. A Monte Carlo of 
100,000 trials was used, drawing from a Gaussian distribution with the same variance as 
the data and using the original sampling, to determine the periodogram power threshold 
level for a 0.1\% false alarm probability (FAP)---i.e., the probability that a 
periodogram peak could be generated by Gaussian noise. Peaks above this level are 
considered significant and were ranked in order of decreasing power. From the MWO time 
series, with a duration of 36.3 years, the top three significant peaks are found at 39.1, 
6.7, and 10.7 years, and the remaining four significant peaks have periods less than 3.5 
years. While the Lomb-Scargle periodogram method is capable of detecting harmonic 
periodicity that extends beyond the duration of the data, such low frequency peaks in a 
more complex time series can be a result of the data window, and must be viewed with 
extreme caution.  From the Keck HIRES-2 time series, with a duration of 14.3 years, no 
significant low frequency periods are detected within the data window, and a peak of 20.5 
years is found beyond it. Finally, from the composite dataset significant peaks are found 
at $9.65\pm0.09$, $7.00\pm0.05$, and $13.4\pm0.2$ years, the first two 
corresponding well to the MWO-alone time series. These results compare well to the 
\citet{Egeland2017} study that combined the MWO data with observations from the Lowell 
Observatory Solar Stellar Spectrograph (SSS), where peaks of 43.3, 6.7, and 12.8 years 
were found. In that study, the long period peak was viewed with skepticism due to a step 
discontinuity in the SSS data corresponding to a CCD upgrade in the SSS. The HIRES-2 data 
does not suffer from such a discontinuity, and the power at low frequencies from the 
composite dataset is diminished, indicating that the $\sim$40 year ``cycle'' indicated by 
the MWO data alone is not supported by these extended observations. Variability on the 
scale of $\sim$7 to $\sim$13 years is the most prominent and reliable, however the 
variations in \lamser are not as clean and Sun-like as the solar-cycle and the term 
``cycle'' for these periodicities should be applied with caution, as discussed more 
generally in \citet{Egeland2017}. The periodogram in Figure \ref{fig4} is expressed in 
units of normalized amplitude $A_N = \sqrt{2 P_N / N}$, where $P_N$ is the usual 
periodogram power normalized by the variance. This normalization is useful for comparing 
time series of different lengths and judging cycle quality as it: (1) is independent of 
the number of observations, (2) has a maximum value of 1, and (3) is a relative measure 
of signal purity for a given peak \citep[see discussion in][]{Egeland2017}.

Individual seasons of the MWO and Keck time series with more than 20 observations were 
analyzed to search for rotational modulation, following previous efforts by 
\citet{Donahue1993} and \citet{Donahue1996}. As with the cycle search, a Lomb-Scargle 
periodogram was employed using a 100,000 trial Monte Carlo to determine the power 
threshold for a 5\% FAP. For a search range between 10 and 40 days, peaks above the 5\% 
FAP threshold are reported as seasonal rotation periods. When significant peaks were 
found, a 100,000 trial periodogram Monte Carlo was used, adjusting the observations 
within their errors to determine the period uncertainty. We compare our results to the 
previous work of \citet{Donahue1993} in Table~\ref{tab2}. Our analysis largely confirms 
the earlier results, which found rotation signals in six seasons ranging from 23.6 to 
28.8 days. We also find significant periods in six seasons, though not the same six found 
by \citeauthor{Donahue1993}, with rotation ranging from 19.4 days to 28.0 days. The 
differences may be ascribed to: (1) \citeauthor{Donahue1993} analyzed individual 
$S$-index observations, while we used a nightly average from typically three observations 
per night, (2) the MWO time series were recalibrated after \citeauthor{Donahue1993}'s 
work, and (3) different observation rejection criteria were employed. No significant 
rotation periods were found in the lower-cadence MWO data beyond the 1993 season, 
which were not analyzed by \citet{Donahue1993}, nor in the Keck HIRES-2 data. The Keck 
observations tend to be clustered around a few dates within a season, making them 
unsuitable for a rotation period search.

To summarize, our analysis of the composite MWO and Keck $S$-index time series indicates 
a mean rotation period of $24.3 \pm 2.7$ days, with strong indications of differential 
rotation, $({\rm max}(P_{\rm rot}) - {\rm min}(P_{\rm rot}))/{\rm min}(P_{\rm rot}) = 
44\%$. Significant long-term variability at 9.7, 7.0, and 13.4 years was found, but it 
does not appear to be dominated by a single period that would indicate a ``clean'' cycle 
as for the Sun. A large amplitude long-period variation of approximately 40 years is 
apparent in the MWO data, but the 54-year composite dataset does not support this 
periodicity being cyclic. Extended uniform data sets are required to determine whether 
such long-period cycles exist in \lamser or other stars.

\subsection{Spectral Energy Distribution}\label{sec2.5} 

As an independent determination of the stellar properties, we performed an analysis of 
the broadband spectral energy distribution (SED) of \lamser and the Gaia DR3 parallax 
\citep[with no systematic offset applied; see, e.g.,][]{StassunTorres:2021}, to derive an 
empirical measurement of the stellar radius, following the procedures described in 
\citet{Stassun:2016,Stassun:2017} and \cite{Stassun:2018}. We adopted the $UBV$ 
magnitudes from \citet{Mermilliod:2006}, the $B_T V_T$ magnitudes from Tycho-2, the 
Str\"omgren $ubvy$ magnitudes from \citet{Paunzen:2015}, the $JHK_S$ magnitudes from 
2MASS, the W1--W4 magnitudes from WISE, the $G_{\rm BP} G_{\rm RP}$ magnitudes from Gaia, 
and the FUV magnitude from GALEX. Together, the available photometry spans the full 
stellar SED over the wavelength range 0.2--22~$\mu$m.

\begin{figure}[t]
\centering\includegraphics[width=\linewidth,trim=95 75 90 90,clip]{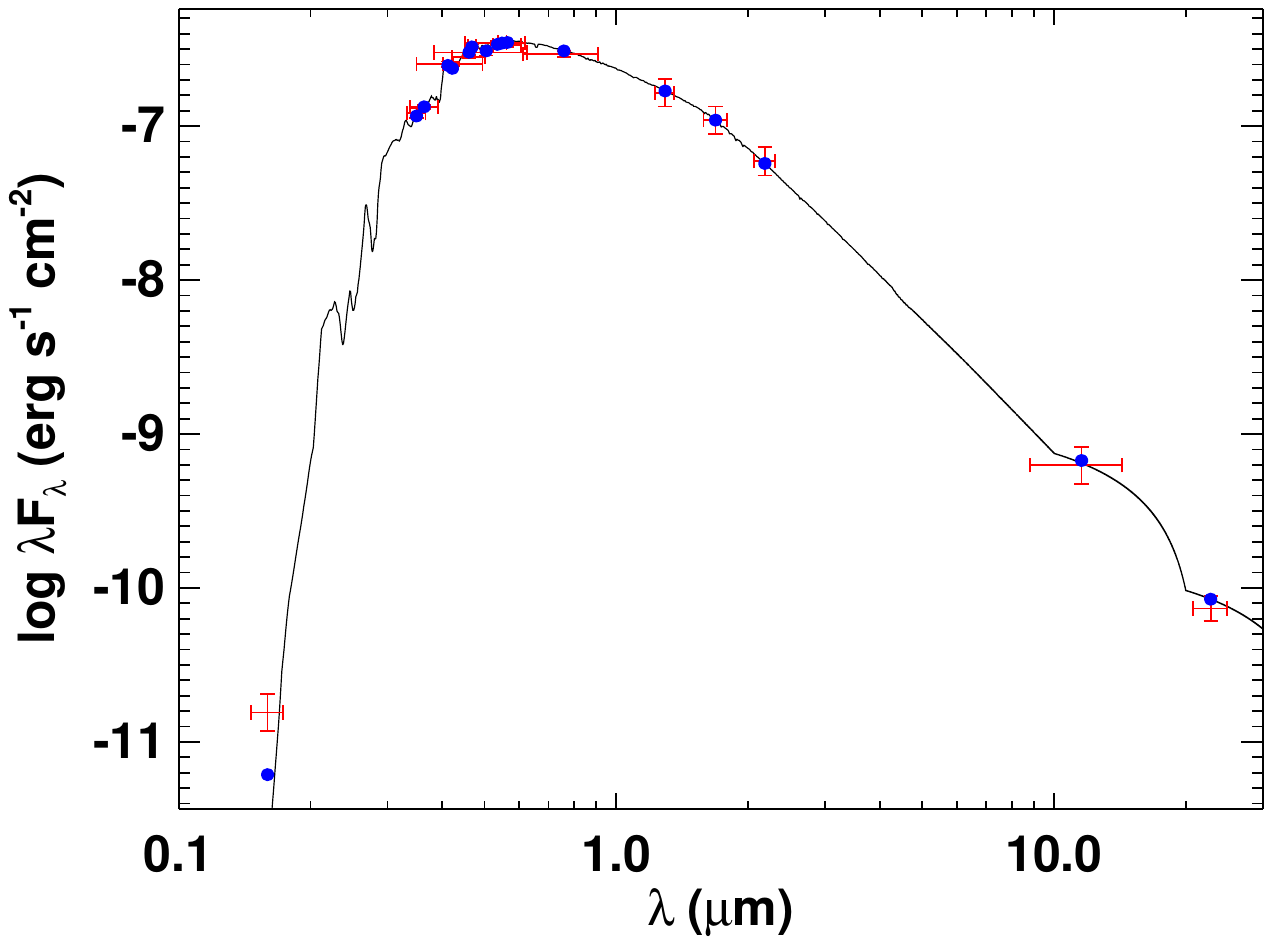}
\caption{Spectral energy distribution of \lamser. Red symbols represent the 
observed photometric measurements, where the horizontal bars represent the effective 
width of the passband. Blue symbols are the model fluxes from the best-fit Kurucz 
atmosphere model (black).\label{fig5}}
\end{figure}

We performed a fit using Kurucz stellar atmosphere models, with the effective temperature 
($T_{\rm eff}$), surface gravity ($\log g$), and metallicity ([M/H]) from 
\cite{Brewer2016}, with uncertainties inflated to account for a realistic systematic 
noise floor. The remaining free parameter is the extinction $A_V$, which we fixed at 
zero due to the star's proximity ($d=11.9$~pc). The resulting fit (Figure~\ref{fig5}) has 
a reduced $\chi^2$ of 1.1, excluding the GALEX FUV flux which indicates a moderate level 
of activity. Integrating the (unreddened) model SED gives the bolometric flux at Earth, 
$F_{\rm bol} = 4.481 \pm 0.052 \times 10^{-7}$ erg~s$^{-1}$~cm$^{-2}$. Taking the 
$F_{\rm bol}$ together with the Gaia parallax gives the bolometric luminosity, 
$L_{\rm bol} = 1.984 \pm 0.023\ L_\odot$, which together with $T_{\rm eff}$ gives 
the stellar radius, $R = 1.349 \pm 0.024\ R_\odot$. In addition, we can estimate 
the stellar mass from the empirical relations of \citet{Torres:2010}, giving $M = 1.17 
\pm 0.07\ M_\odot$, which is consistent with that obtained directly from R and $\log g$ 
($M = 1.11 \pm 0.12\ M_\odot$). These estimates of the radius and mass can be 
compared to the adopted values from asteroseismology in Section~\ref{sec3.1}.

\subsection{Hubble Space Telescope Data}\label{sec2.6}

\begin{deluxetable}{lcc}
\setlength{\tabcolsep}{10pt}
\tablecaption{Measured FUV Line Fluxes for \lamser \label{tab3}}
\tablehead{ \colhead{Ion(s)} & \colhead{Wavelength} & \colhead{Flux at Earth} \\  
\colhead{}  &  \colhead{[\AA]}  & \colhead{[10$^{-15}$ ergs cm$^{-2}$ s$^{-1}$]}}
\startdata
C {\sc iii}$^1$            & 1175   & 29.3  $\pm$ 1.3     \\ 
Si {\sc iii}$^1$           & 1198   & 0.27  $\pm$ 0.19    \\
O {\sc i}$^1$              & 1304   & 54.8  $\pm$ 1.1     \\
C {\sc ii}$^1$             & 1335   & 71.3  $\pm$ 0.9     \\   
Cl {\sc i}$^2$             & 1351.7 & 2.2   $\pm$ 0.3  \\  
Si {\sc iv}                & 1393.8 & 27.3  $\pm$ 0.7     \\  
O {\sc iv}                 & 1399.8 & 1.3   $\pm$ 0.3     \\  
O {\sc iv}                 & 1401.2 & 1.9   $\pm$ 0.3     \\  
Si {\sc iv}$^2$            & 1402.8 & 12.44 $\pm$ 0.6  \\   
Si {\sc iv}+O {\sc iv}$^2$ & 1404.8 & 2.40  $\pm$ 0.6  \\  
O {\sc iv}$^2$             & 1407.4 & 0.30  $\pm$ 0.16 \\  
N {\sc iv}                 & 1486.3 & 1.2   $\pm$ 0.3     \\ 
Continuum$^3$              & 1506   & 21.4  $\pm$ 0.8     \\  
Si {\sc ii}                & 1526.5 & 8.9   $\pm$ 0.9     \\   
Si {\sc ii}                & 1533.7 & 8.4   $\pm$ 0.8     \\   
C {\sc iv}$^2$             & 1548   & 40.2  $\pm$ 1.3  \\ 
C {\sc iv}$^2$             & 1550   & 18.2  $\pm$ 1.1  \\ 
C {\sc i}$^{1,2}$          & 1561   & 13.5  $\pm$ 1.0  \\  
He {\sc ii}                & 1640.7 & 17.5  $\pm$ 1.5     \\  
C {\sc i}$^1$              & 1657   & 62.7  $\pm$ 2.5     \\  
O {\sc iii}                & 1666.2 & 2.4   $\pm$ 1.2     \\   
Si {\sc ii}                & 1808.0 & 26.9  $\pm$ 3.1     \\
Si {\sc ii}                & 1817.1 & 138   $\pm$ 5       \\   
Al {\sc iii}               & 1854.6 & 10.3  $\pm$ 3.6     \\
Si {\sc iii}               & 1892.0 & 88.4  $\pm$ 6.6     \\  
C {\sc iii}                & 1908.7 & 50.8  $\pm$ 7.9     \\  
\enddata
\tablecomments{All fluxes are from direct integration, except
  $^1$Multiple lines combined, $^2$Voigt function fitting and deblending,
  $^3\pm$5~\AA~integration of a largely line-free region.}
\vspace*{-24pt}
\end{deluxetable}
\vspace*{-12pt}

As a probe of the physical environment between the photosphere and the corona, we also 
observed \lamser using the Cosmic Origins Spectrograph (COS) on the Hubble Space 
Telescope (HST). We obtained a low resolution G140L FUV spectrum with an integration time 
of 1976~s on 2020 September 2 (program 15991), which was reduced with standard pipeline 
processing. We computed integrated fluxes for isolated emission lines in the rest 
frame of the star above nearby pseudo-continua, together with RMS errors. We estimated 
continua from linear fits to clusters of low points on either side of the line in 
question.  When the target line was blended, two methods were employed. Some weaker 
blends were removed by fitting a Voigt function to better mimic the convolved 
instrumental profile and line shape, leaving the residual target line for flux 
integration as before. In some cases, we fit the entire complex of lines with multiple 
Voigt functions. The results are shown in Table~\ref{tab3}.  Tests on isolated lines 
demonstrated that straight integration and Voigt fitting yielded similar results, 
typically within $\pm$5\%. In several cases, multiple nearby lines of the same ion were 
combined. Following \citet{Ayres2020}, we also measured a 10~\AA\ segment of relatively 
line-free FUV pseudo-continuum centered at 1506~\AA.

\begin{deluxetable}{lcccc}
\tablecaption{Comparison of FUV Surface Fluxes \label{tab4}}
\tablehead{\colhead{Ion(s)} & \colhead{Wavelength} & \multicolumn{3}{c}{Surface Flux [10$^{3}$ ergs cm$^{-2}$ s$^{-1}$]} \\
\colhead{}  &  \colhead{[\AA]}  &  \lamser & Sun$^2$ & $\alpha$ Cen A$^2$ }
\startdata
C {\sc iii}$^1$ & 1175   & 4400  $\pm$ 200 & 2250 & $\cdots$           \\
O {\sc i}$^1$   & 1304   & 8230  $\pm$ 160 & 5490 & 5800 $\pm$ 290     \\
C {\sc ii}$^1$  & 1335   & 10700 $\pm$ 140 & 7000 & 7000  $\pm$ 350    \\
Cl {\sc i}      & 1351.7 & 330   $\pm$ 43  &  252 & $\cdots$           \\
Si {\sc iv}     & 1393.8 & 4110  $\pm$ 104 & 1690 & 3200 $\pm$ 160$^1$ \\
Si {\sc iv}     & 1402.8 & 1870  $\pm$ 86  &  875 & $\cdots$           \\
Continuum       & 1506   & 3210  $\pm$ 117 & 1780 & $\cdots$           \\
C {\sc iv}      & 1548   & 6030  $\pm$ 198 & 3800 & 6200 $\pm$ 310$^1$ \\
C {\sc iv}      & 1550   & 2730  $\pm$ 171 & 1960 & $\cdots$           \\
\enddata
\tablecomments{$^1$Multiple lines combined, $^2$Results from \citet{Ayres2020}.} 
\end{deluxetable}
\vspace*{-12pt}

We can compare the FUV fluxes of \lamser to the Sun and to $\alpha$~Cen~A, a somewhat 
older \citep[5.3~Gyr;][]{Joyce2018}, more metal rich \citep[Fe/H=+0.24;][]{Morel2018} 
solar analog (see Table~\ref{tab4}).  Lower chromospheric surface fluxes in Cl~{\sc i} 
and O~{\sc i} (with temperatures of peak emissivity $\log T_{\rm peak}$ = 3.8--3.9) are 
1.3 to 1.5 times larger in \lamser; the upper chromospheric C~{\sc ii} flux ($\log T_{\rm 
peak} = 4.5$) is similarly $\approx 1.5\times$ enhanced. Moving to lines formed in the 
stellar transition region, C~{\sc iii} ($\log T_{\rm peak} = 4.8$) is $\approx2\times$ 
enhanced in \lamser relative to the Sun, Si~{\sc iv} ($\log T_{\rm peak}$ = 4.9) is 
further enhanced, at a factor of 2.3 and 1.9 relative to the Sun and $\alpha$~Cen~A, 
respectively (the higher metallicity of $\alpha$~Cen~A may boost its emission). In the 
C~{\sc iv} doublet ($\log T_{\rm peak}$ = 5.0), the enhancement in \lamser returns to a 
factor of $\approx$1.5 (here some optical depth effects may play a role).  These results 
are generally in line with the more active corona of \lamser---which shows 3.8$\times$ 
the solar surface $F_X$---and with the reduced activity enhancements relative to the 
corona that are expected for lower $T_{\rm peak}$ emission in the chromosphere and 
transition region \citep[e.g.,][among many]{Ayres2022}.

Several density-sensitive line ratios are available in the HST spectra.  We use the 
intersection of these results to estimate the electron density in the transition region.  
The Si~{\sc iii}(1892~\AA)/C~{\sc iii}(1909~\AA) ratio, with $\log T_{\rm peak} \sim 
4.7$, yields $\log n_e = 9.87_{-0.18}^{+0.08}$ using \citet{Keenan1987}.  The ratio 
C~{\sc iii}(1908~\AA)/Si~{\sc iv}(1402~\AA), with $\log T_{\rm peak} \sim 4.8$, gives 
$\log n_e = 9.98_{-0.06}^{+0.05}$. The ratios O~{\sc iii}(1666~\AA)/Si~{\sc iv}(1402~\AA) 
and C~{\sc iii}(1908~\AA)/O~{\sc iii}(1666~\AA), also with $\log T_{\rm peak} \sim 4.8$, 
were less certain, affected by larger errors in the O~{\sc iii} line; these yield $\log 
n_e = 10.80_{-0.41}^{+0.40}$ and $9.76_{-1.44}^{+0.30}$, respectively \citep[using 
][]{Keenan1988}.  Another $\log T_{\rm peak} \sim 4.8$ diagnostic, C~{\sc 
iii}(1908~\AA)/Al~{\sc iii}(1863~\AA), implies $\log n_e = 10.24_{-0.21}^{+0.21}$ 
\citep[employing ][]{Keenan1990}. A hotter diagnostic, the ratio O~{\sc 
iv}(1401~\AA)/O~{\sc iv}(1407~\AA) at $\log T_{\rm peak} \sim 5.1$ \citep[using results 
of][]{Brage1996} unfortunately gives only a weak limit of $\log n_e < 10.3$ due to large 
errors on the fluxes. Combining the cooler diagnostics, we find an average $\langle \log 
n_e \rangle = 10.06_{-0.06}^{+0.05}$ at $\log T_{\rm peak} \sim 4.8$. For comparison, 
$\langle \log n_e \rangle = 10.0$ using the hotter O~{\sc iv} ratio in the Sun 
\citep[e.g., ][]{Rao2022}.  This suggests that \lamser has transition region densities 
similar to, or perhaps slightly lower than, the Sun at fixed temperature, which is 
consistent with its slightly lower surface gravity.  We caution, however, that 
assumptions intrinsic to the line ratio method make the results uncertain \citep[see 
discussion in][]{Judge2020}.

In summary, \lamser has chromospheric and transition region fluxes broadly consistent 
with a star which is somewhat more coronally active, and has slightly lower density than 
the Sun.

\section{Interpretation}\label{sec3}

\subsection{Asteroseismic Modeling}\label{sec3.1}

Using the oscillation frequencies listed in Table~\ref{tab1}, spectroscopic constraints 
on $T_{\rm eff}$ and $\mathrm{[M/H]}$ from \cite{Brewer2016}, and the luminosity from 
Section~\ref{sec2.5}, five teams attempted to infer the properties of \lamser from 
asteroseismic modeling. A variety of stellar evolution codes and fitting methods were 
employed, including ASTEC (AMP) \citep{JCD08, Metcalfe2009}, GARSTEC (BASTA) 
\citep{weiss08, Aguirre2022}, MESA \citep{Paxton2015, liyg23}, and YREC 
\citep{demarque_yrec_2008}. We found reasonable agreement between the results for the 
stellar radius and mass, with individual estimates ranging from $R=1.33$--1.38~$R_\odot$ 
and $M=1.03$--1.13~$M_\odot$, but there was a significant spread in stellar age with 
inferences between 5.4--8.6~Gyr around a median value of $7.0\pm0.8$~Gyr. For consistency 
with the rotational evolution modeling in Section~\ref{sec3.3}, we adopted the modeling 
results from YREC, which yielded the median estimates of radius and mass with an age at 
the young end of the distribution (see Table~\ref{tab5}).

\begin{deluxetable}{lcc}
  \setlength{\tabcolsep}{14pt}
  \tablecaption{Adopted Properties of the Exoplanet Host Star \lamser\label{tab5}}
  \tablehead{\colhead{}            & \colhead{\lamser}& \colhead{Source}}
  \startdata
  $T_{\rm eff}$ (K)                & $5901 \pm 78$          & 1 \\
  $[$M/H$]$ (dex)                  & $+0.04 \pm 0.07$       & 1 \\
  $\log g$ (dex)                   & $4.22 \pm 0.08$        & 1 \\
  $B-V$ (mag)                      & $0.60$                 & 2 \\
  $\log R'_{\rm HK}$ (dex)         & $-5.004$               & 2 \\
  $P_{\rm rot}$ (days)             & $24.3^{+3.7}_{-4.9}$   & 3 \\
  $|B_{\rm d}|$ (G)                & $2.73, 2.12$           & 4 \\
  $|B_{\rm q}|$ (G)                & $1.90, 2.21$           & 4 \\
  $|B_{\rm o}|$ (G)                & $1.37, 2.44$           & 4 \\
  $L_X$ ($10^{27}$~erg~s$^{-1}$)   & $6.8 \pm 0.6$          & 5 \\
  Mass-loss rate ($\dot{M}_\odot$) & $1.6 \pm 0.2$          & 5 \\
  Luminosity ($L_\odot$)           & $1.984 \pm 0.023$      & 6 \\
  Mass ($M_\odot$)                 & $1.09 \pm 0.04$        & 7 \\
  Radius ($R_\odot$)               & $1.363 \pm 0.031$      & 7 \\
  Age (Gyr)                        & $5.4 \pm 0.7$          & 7 \\
  \hline
  Torque ($10^{30}$~erg)           & $2.01^{+0.82}_{-0.64}$ & 8 \\
  \enddata
  \tablerefs{(1)~\cite{Brewer2016}; (2)~\cite{Baliunas1996}; (3)~Section\,\ref{sec2.4}; 
    (4)~Two ZDI reconstructions in Section\,\ref{sec3.2}; (5)~Section\,\ref{sec2.3};
    (6)~Section\,\ref{sec2.5}; (7)~Section\,\ref{sec3.1}; (8)~Section\,\ref{sec3.4}}
\vspace*{-24pt}
\end{deluxetable}
\vspace*{-12pt}

The YREC results were obtained from a grid of models that were constructed with the Yale 
Stellar Evolution Code \citep{demarque_yrec_2008}. All models were constructed with the 
same microphysics inputs---OPAL opacities \citep{1996ApJ...464..943I} supplemented with 
low temperature opacities from \citet{2005ApJ...623..585F}, the OPAL equation of state 
\citep{nayfonov}, nuclear reaction rates from \citet{1998RvMP...70.1265A}, except for the 
$^{14}$N$(p,\gamma)^{15}$O reaction, for which we use the rate of 
\citet{2004PhLB..591...61F}.  Additionally, models included gravitational settling of 
helium and heavy elements using the formulation of \citet{thoul_element_1994}, with the 
diffusion coefficient modified using the mass dependent factor of 
\citet{viani_diffusion}.

We first determined mass and $\log g$ from the global asteroseismic parameters using the 
Yale-Birmingham pipeline \citep[YB;][]{YB}. This step informed us of the mass range---we 
used the inferred mass and a $\pm 3\sigma$ range around it, i.e., 0.99 M$_\odot$ to 1.25 
M$_\odot$ to construct a grid of models. For each mass, models were created with seven 
values of the mixing length parameter spanning $\alpha_{\rm MLT}=1.5$ to 2.3, and initial 
helium abundances from 0.20 to 0.32. The initial [M/H] of the models spanned the range 
0.0 to +0.3 dex to account for the diffusion and settling of heavy elements. The models 
were evolved from the zero-age main sequence (ZAMS). Models along a track were output 
within $\pm 3\sigma$ of the $\log g$ returned by the YB pipeline and their frequencies 
calculated with the code of \citet{antia}.

The properties of \lamser were determined as follows.  We first corrected for the surface 
term using the two-term correction proposed by \citet{ball_correction_2014}. The 
corrected frequencies were used to define a $\chi^2$:
\begin{equation}
\chi^2(\nu)=\frac{1}{N-1}\sum_{nl}\frac{(\nu_{nl}^{\rm obs}-\nu_{nl}^{\rm corr})^2}{\sigma^{\rm obs}_{nl}},
\label{eq:eqchinu}
\end{equation}
$N$ being the number of modes, which was then used to determine a likelihood function:
\begin{equation}
{\mathcal L}(\nu)=C\exp\left(-\frac{\chi^2(\nu)}{2}\right),
\label{eq:nulike}
\end{equation}
$C$ being the normalization constant. We also defined a likelihood for each of the other 
observables, $T_{\rm eff}$, [M/H] and luminosity $L$. For instance, the likelihood for 
effective temperature was defined as
\begin{equation}
{\mathcal L}(T_{\rm eff})=D\exp(-\chi^2(T_{\rm eff})/2),
\label{eq:tcal}
\end{equation}
with
\begin{equation}
\chi^2(T_{\rm eff})=\frac{(T^{\rm obs}_{\rm eff}-T^{\rm model}_{\rm eff})^2}{\sigma^2_{ T}},
\label{eq:chit}
\end{equation}
where $\sigma_{T}$ is the uncertainty on the effective temperature, and $D$ the constant 
of normalization. We similarly defined the likelihoods for [M/H] and $L$. The total 
likelihood for each model is then
\begin{equation}
{\mathcal L}_{\rm total}=C{\mathcal A}{\mathcal L}(\nu){\mathcal L}(T_{\rm eff}){\mathcal L}([M/H]){\mathcal L}(L).
\label{eq:totlike}
\end{equation}
The quantity ${\mathcal A}$ is a prior that we used to down-select models with ages 
greater than 13.8 Gyr; without this prior the likelihood distribution would have a sharp 
cut-off. We define ${\mathcal A}$ as
\begin{equation}
{\mathcal A}=
\begin{cases}
1, & \hbox{if}\;\; \tau <=13.8\;\hbox{Gyr}\\
\exp\left[-\frac{(13.8-\tau)^2}{2\sigma^2_\tau}\right] & \hbox{otherwise},\\
\end{cases}
\label{eq:wage}
\end{equation}
where the age $\tau$ is in units of Gyr, and  $\sigma_{\tau}$ is chosen to be 0.1 Gyr.  

The medians of the marginalized likelihoods of the ensemble of models were used to 
determine the stellar properties, after converting them to a probability density by 
normalizing the likelihood by the prior distribution of the property. We repeated the 
exercise by perturbing each of the non-seismic inputs ($T_{\rm eff}$, $[M/H]$, and $L$) 
by a normally distributed random amount with variance given by the observational errors. 
The distribution given by the ensemble of medians was used to determine the final stellar 
properties shown in Table~\ref{tab5}.

\subsection{Zeeman Doppler Imaging}\label{sec3.2} 

Like the PEPSI data presented in Section~\ref{sec2.2}, polarized Zeeman signatures in the 
individual lines of reduced Neo-NARVAL spectra are dominated by noise. We employed the 
LSD method to extract a cross-correlation line profile from a list of photospheric lines 
\citep{Donati1997, Kochukhov2010}. The line mask was chosen to be closest to the 
fundamental parameters of \lamser in the grid of \cite{Marsden2014}, keeping lines deeper 
than 40\% of the continuum with no telluric contamination. The normalized Land\'e 
factor of LSD profiles is close to 1.2, while their normalized wavelength is equal to 
650~nm.

\begin{figure}[t]
\centering\includegraphics[width=\linewidth,trim=10 10 0 0,clip]{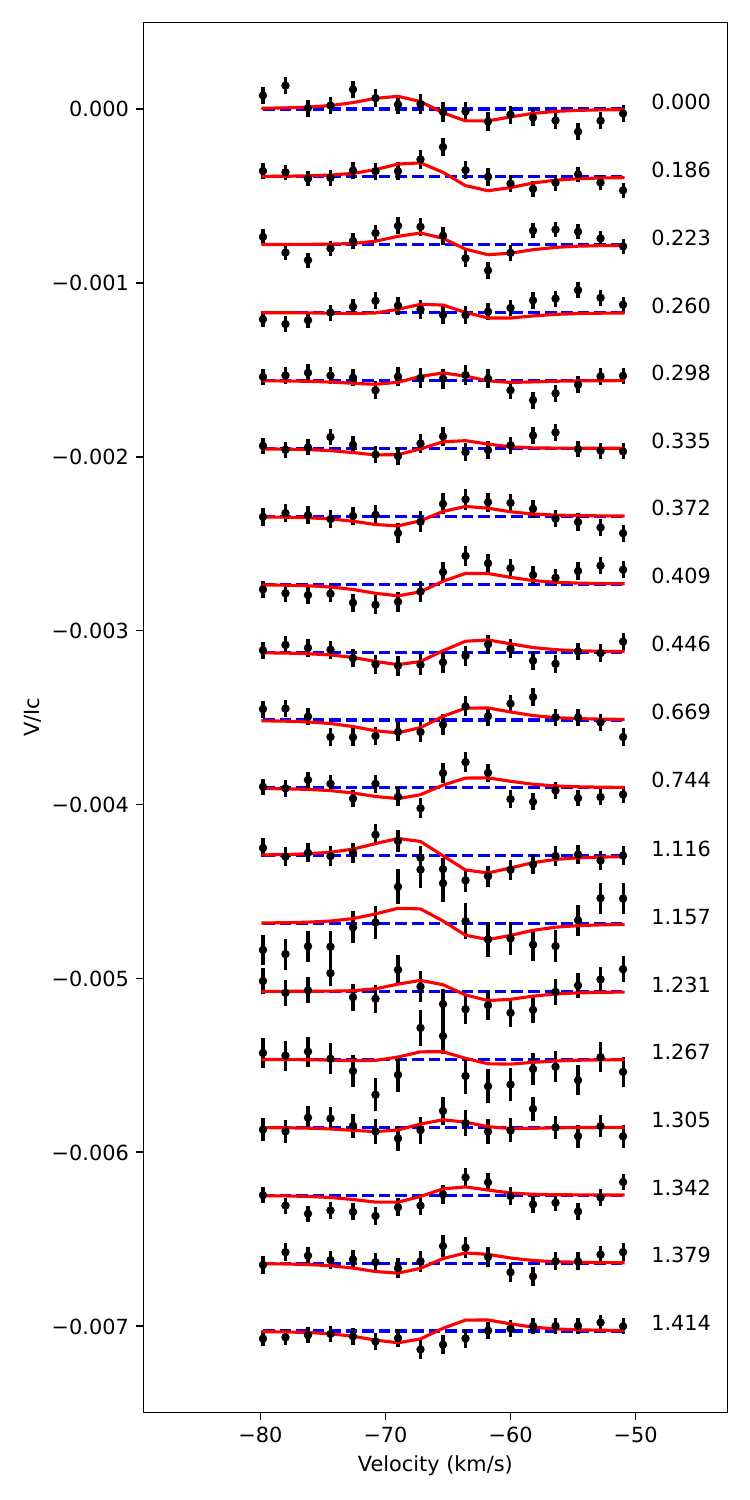}
\caption{Stokes~$V$ LSD profiles obtained with Neo-NARVAL (black dots), and synthetic 
profiles produced by ZDI (red curves). Successive observations are vertically shifted for 
clarity, and the rotational phase $\Phi$ of the observation is listed along the right 
side.\label{fig6}}
\end{figure}

Owing to the lack of lines bluer than 470~nm, we ended up with slightly less than 1,700 
available spectral lines and a S/N of LSD profiles (per 1.8~km~s$^{-1}$ velocity bin) 
between 11,000 and 24,000, with a mean value of 19,000. Even after the LSD processing, 
Zeeman signatures remained barely visible by eye, which is consistent with the small 
polarized amplitude found by PEPSI (see Section~\ref{sec2.2}). Applying the criterion of 
\cite{Donati1992, Donati1997}, we obtain only four marginal detections (false alarm 
probability between $10^{-4}$ and $10^{-5}$), with all other observations considered 
non-detections. The LSD profiles are shown in Figure~\ref{fig6}.

Running ZDI \citep{Semel1989} on polarization signatures close to the detection threshold 
is not ideal, but previous studies have shown that it is possible to reconstruct magnetic 
maps even when the signatures are dominated by noise, provided that a sufficient number 
of observations are combined together in the inversion process \citep{Petit2010, 
Petit2022}. Here, we use the ZDI implementation of \cite{Folsom2018a, Folsom2018b}, in a 
procedure closely following the one presented by \cite{Petit2021}. The surface magnetic 
field is described by the set of spherical harmonic equations in \cite{Donati2006}, and 
we limited the expansion to $\ell_{\rm max} = 10$ because the magnetic model was not 
improved by including higher-order terms.

\begin{figure}[t]
\centering\includegraphics[width=\linewidth,trim=35 10 30 10,clip]{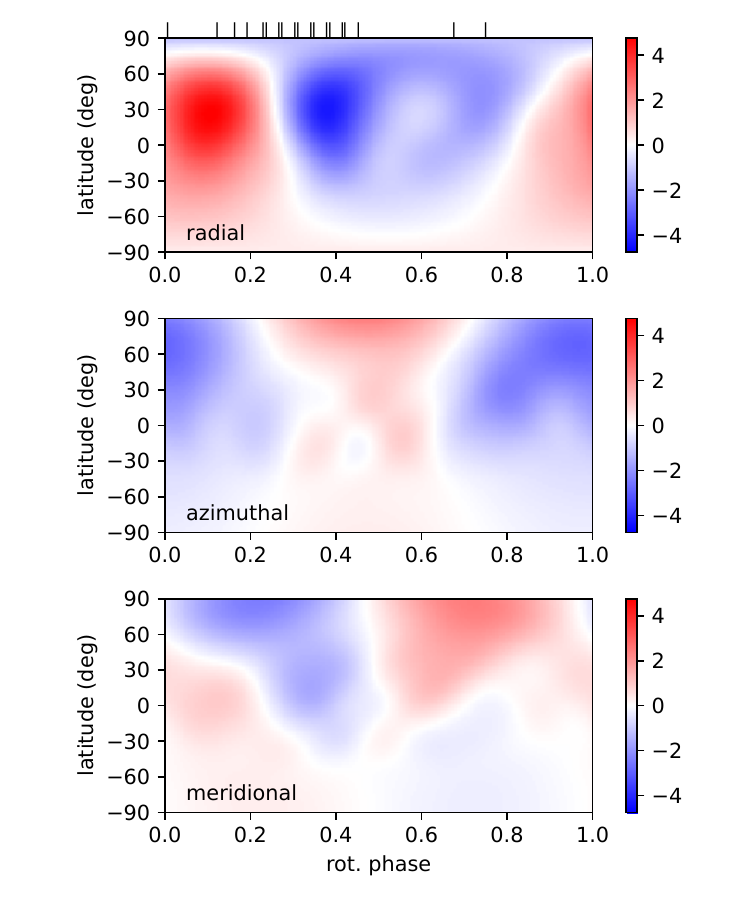}
\caption{Large-scale surface magnetic geometry of \lamser. The three panels display 
the field components in spherical coordinates, adopting an equatorial projection. Color 
bars on the right show the field strength in Gauss. Vertical ticks above the top panel 
indicate observed rotational phases. Latitudes below $-40^\circ$ are not observed.
\label{fig7}}
\end{figure}

We adopted a projected rotational velocity of 2~km~s$^{-1}$ \citep{Brewer2016}, resulting 
in an inclination angle of 50$^\circ$ when combined with our estimates of $P_{\rm rot}$ 
and $R$. The radial velocity of our LSD profiles was $-$65.9~km~s$^{-1}$, which is about 
0.5~km~s$^{-1}$ larger than recent estimates \citep{Soubiran2018}. Following 
\cite{Petit2008}, we ran a series of ZDI inversions assuming different values of the 
rotation period and found that the best fit was obtained for $P_{\rm rot} = 26.87$~d. 
Following this estimate, the rotational phase $\Phi$ of each observation was calculated 
using the following ephemeris:
\begin{equation} \label{equ:emp} 
\mathit{HJD}_{\mathrm{obs}} = \mathit{HJD}_{0} + P_{\rm{rot}} \times \Phi 
\end{equation} 
where the initial Heliocentric Julian date $\mathit{HJD}_{0}$ was from our first 
observation at 2459406.373. It is clear from Figure~\ref{fig6} that the phase coverage is 
very good between $\Phi = 0$ and 0.4 (observed over two consecutive rotation cycles), 
while only two observations have phases above 0.5. We also performed a search for 
differential rotation following the method of \cite{Petit2002} but failed to measure a 
surface shear. The target reduced $\chi^2$ of the ZDI inversion was fixed at 1.075, 
because adopting lower values led to clear signs of overfitting (visible as a sharp 
increase in the average field strength and field complexity). The resulting map is shown 
in Figure~\ref{fig7}, while the synthetic Stokes~$V$ LSD profiles are plotted with red 
lines in Figure~\ref{fig6}.

The reconstructed magnetic geometry has an average field strength of 2~G, with a maximum 
local peak strength of 5~G. A majority of the magnetic energy (87\%) shows up in the 
poloidal field component, and more specifically in the dipole component that hosts 71\% 
of the poloidal magnetic energy. The dipole strength is equal to 2.9~G, and it is very 
inclined with respect to the spin axis, with a negative pole located at latitude 
$\sim$11$^\circ$. Unsurprisingly, this very non-axisymmetric magnetic configuration 
leads to only 9\% of the magnetic energy in modes with $m = 0$.

Considering the low amplitude of polarization signatures in the Neo-NARVAL LSD profiles, 
we carried out an independent ZDI reconstruction with an alternative inversion code 
\citep{Kochukhov2014, Rosen2016, Lehtinen2022}. This inversion adopted the same $P_{\rm 
rot}$, $\ell_{\rm max}$, $i$, and $v_{\rm e}\sin i$, resulting in a qualitatively similar 
magnetic field distribution to the one illustrated in Figure~\ref{fig7} but with a 
somewhat stronger and more structured magnetic field map. In this case, we found an 
average field strength of 3.7~G and a maximum local strength of 8.9~G. Contributions of 
the poloidal and toroidal components are nearly equal, with the dipole component 
containing 48\% of the total magnetic field energy and 31\% of the poloidal field energy. 
The discrepancies of these parameters with the outcome of the first ZDI reconstruction 
likely reflect intrinsic limitations of ZDI based on low S/N data. Nevertheless, the 
dipole field characteristics (strength 2.1~G, obliquity 98\degr\ towards the positive 
pole) are similar to those obtained in the first inversion.

To estimate the rate of angular momentum loss for \lamser, we use the braking law of 
\citet{FinleyMatt2018} (see Section \ref{sec3.4}). This braking law requires the polar 
strengths of the dipole, quadrupole and octupole components of stellar magnetic field as 
inputs. These can be obtained from the reconstructed ZDI maps. However, the 
magnetohydrodynamic simulations used to construct the \citet{FinleyMatt2018} braking law 
were run using only axisymmetric magnetic field modes whereas the ZDI maps contain both 
axisymmetric and non-axisymmetric components. In order to calculate the equivalent polar 
field strengths needed for the braking law from each ZDI map, we used the method we 
employed in \citet{Metcalfe2022}. Briefly, this method calculates the magnetic flux in 
each of the dipole, quadrupole and octupole components of the ZDI map, i.e.\ accounting 
for both the axisymmetric and non-axisymmetric components. We then determine the polar 
field strengths of a purely axisymmetric dipole, quadrupole and octupole that reproduces 
the respective magnetic fluxes of each component from the ZDI map. These are the field 
strengths reported in Table~\ref{tab5} (first, second reconstruction) and used in 
the braking law.

\subsection{Rotational Evolution}\label{sec3.3}

We fit a rotational evolution model to \lamser following the methodology described in 
\citet{Metcalfe2020}. We use slightly different values for two braking law parameters: a 
braking normalization of $f_k= 8.53$ for the standard law and $f_k = 8.97$ and Ro$_{\rm 
crit} = 2.01$ for the weakened magnetic braking law, derived from calibrating the braking 
law to the asteroseismic rotator sample of \citet{Hall2021}, open clusters, and the Sun 
(Saunders et al. in prep). This amounts to a braking normalization that is $\sim$40\% 
higher and Ro$_{\rm crit}$ that is 7\% lower than that used in \citet{Metcalfe2020} for 
the weakened law, and a braking normalization $\sim$25\% higher for the standard law. 
These changes would tend to make a star of a given age and mass spin more slowly, 
although weakened magnetic braking occurs at slightly faster rotation rates.

We search for a best-fit model that matches our observed surface temperature, luminosity, 
and surface metallicity, with asteroseismic priors on the mass, age, and mixing length as 
described in \citet{Metcalfe2020}. Our best-fit model reproduces all surface observables 
and priors (with the exception of rotation) within $1\sigma$. For a standard model, we 
predict a rotation period of $34 \pm 6$~days, while weakened magnetic braking predicts a 
period of $18 \pm 2$~days. The weakened magnetic braking model is consistent with the 
most rapid seasonal rotation rate observed for \lamser. If \lamser is in fact viewed at a 
moderate inclination and has solar-like differential rotation, we might expect the 
observed rotation period to be slower than the equatorial rotation period that is 
predicted by the solid body stellar models. However, its mean rotation period is in mild 
tension with both the weakened magnetic braking and standard case, and does not 
conclusively distinguish between the two scenarios.

\subsection{Magnetic Evolution}\label{sec3.4}

Bringing together the magnetic field properties derived in Section~\ref{sec3.2}, the 
mass-loss rate estimated in Section~\ref{sec2.3}, the range of rotation periods measured 
in Section~\ref{sec2.4}, and the asteroseismic radius and mass from Section~\ref{sec3.1}, 
we can use the prescription of \cite{FinleyMatt2018} to estimate the wind braking torque 
of \lamser. We repeat the calculation using the magnetic field properties ($B_d, B_q, 
B_o$) from two independent ZDI reconstructions that relied on the same set of Stokes~$V$ 
profiles (see Section~\ref{sec3.2}). In Table~\ref{tab5} we report the average torque 
resulting from the two calculations, and we adopt half of the difference between them as 
the uncertainty arising from the magnetic field properties (11\%). The total uncertainty 
on the torque includes additional contributions from the rotation period (6\%), mass-loss 
rate (4\%), radius (4\%), and mass (1\%), and it reflects the range of possible torques 
when all quantities are shifted by $\pm 1\sigma$.

\begin{figure}[t]
\centering\includegraphics[width=\linewidth]{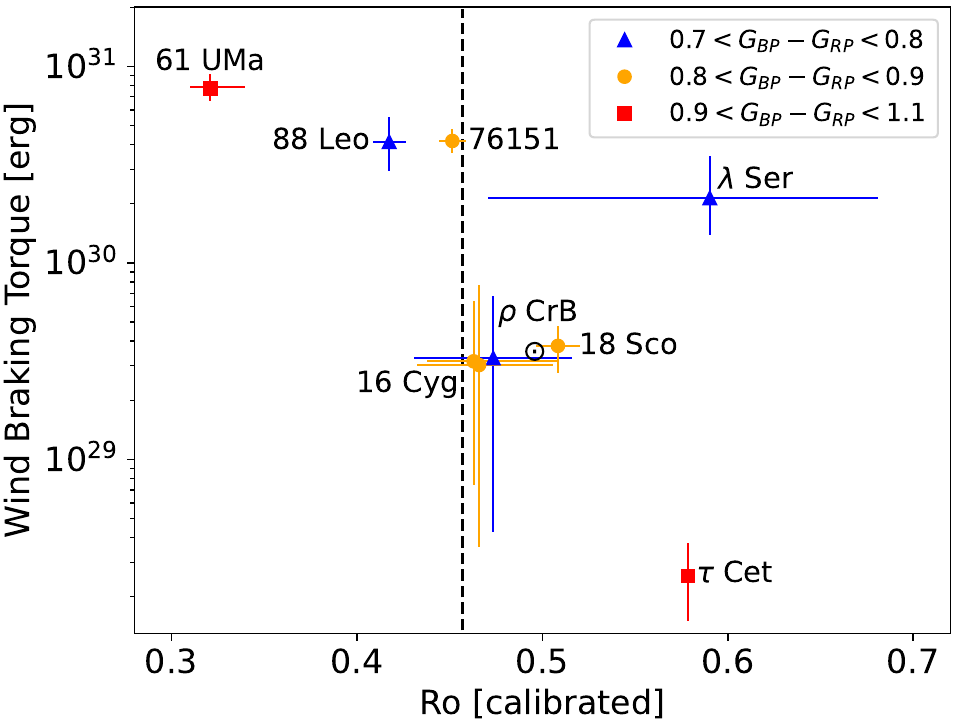}
\caption{Evolution of the wind braking torque with Ro from the calibration of 
\cite{Corsaro2021}. Points are grouped by Gaia color, corresponding to solar analogs 
(yellow circles), and hotter (blue triangles) or cooler stars (red squares). The 
empirical constraint for Ro$_{\rm crit}$ on this scale is shown with a vertical dashed 
line.\label{fig8}}
\end{figure}

The estimated wind braking torque for \lamser is shown relative to several other stars in 
Figure~\ref{fig8}. We calculate the Rossby number for each star from the Gaia 
$G_{BP}-G_{RP}$ color, using the asteroseismic calibration from \cite{Corsaro2021}. We 
estimate the wind braking torque following the methodology outlined in 
\cite{Metcalfe2021, Metcalfe2022}, while the solar point ($\odot$) comes from 
\cite{Finley2018}. On this scale, the empirical value of the Rossby number that 
corresponds to the onset of weakened magnetic braking is ${\rm Ro_{crit}}=0.46$ (dashed 
line). The horizontal error bar for \lamser corresponds to the range of seasonal rotation 
periods identified in Section~\ref{sec2.4}, while the vertical error bar is dominated by 
uncertainties in the strength and morphology of the large-scale magnetic field (see 
Section~\ref{sec3.2}), with progressively smaller contributions from the rotation period, 
mass-loss rate, radius, and mass. Even considering the uncertainties, the wind braking 
torque for \lamser is much higher than for other stars with similar Ro \citep[cf.\ 
$\tau$~Cet;][]{Metcalfe2023}.

The relatively high wind braking torque for \lamser cannot easily be attributed to an 
erroneous measurement. Our previous LBT observations of flat activity stars ($\rho$~CrB, 
16~Cyg~A\,\&\,B) resulted in null detections, while the Stokes~$V$ signature for \lamser 
is strong and consistent with the lower S/N measurements from TBL. There is considerable 
scatter in the \cite{Wood2021} relation between X-ray flux and mass-loss rate, but the 
two subgiants in the calibration both have {\it higher} mass-loss rates than predicted 
from their X-ray flux. Despite the low mean activity level, the rotation rate inferred 
from MWO observations appears consistently in multiple seasons and across the complete 
data set. The radius and mass inferred from asteroseismology are both precise, and they 
agree with the independent estimates from the SED in Section~\ref{sec2.5}. The difficulty 
of matching the stellar properties with rotational evolution models that assume either 
standard spin-down or weakened magnetic braking also suggests that \lamser may have taken 
an unusual path to its present configuration. The asteroseismic age is consistent with 
the activity-age relation for solar analogs \citep{huber22}, but the remaining 
uncertainty prevents an unambiguous interpretation of the other stellar properties.

\section{Discussion}\label{sec4}

Our data on \lamser adds an interesting piece to our understanding of rotation, 
magnetism, and dynamos in old sun-like stars. Stars more active than the Sun spin down as 
they age, so it is natural to extrapolate this behavior to older and less active stars. 
However, it is not surprising that our intuition, developed in a limited empirical 
domain, would break down in the face of the time domain revolution in stellar 
astrophysics. The first hint was the lack of very slowly rotating stars in the 
groundbreaking \cite{mcquillan14} Kepler sample. The Sun, a median-aged disk star, was at 
the upper end of the observed distribution of stellar rotation periods on the main 
sequence. However, this could be induced either by a true cessation of spin-down or a 
threshold in detectability \citep{vansaders19}. A much stronger indication of disrupted 
magnetic braking was the discovery of counter-examples: stars rotating too rapidly to 
have experienced the degree of magnetic braking predicted by standard models. The 
observed pattern favored a dramatic decrease in the efficacy of magnetized winds above a 
critical Rossby threshold \citep{vanSaders2016}. This phenomenon requires a transition 
such that inactive stars experience minimal angular momentum losses over long timescales. 
It does not directly test the origin of this transition, or the degree to which it is 
sudden rather than gradual.

In a series of papers \citep{Metcalfe2021, Metcalfe2022, Metcalfe2023} we have mapped out 
the magnetic field strength and morphology of stars close to the disrupted magnetic 
braking threshold, and used these data to infer integrated instantaneous angular momentum 
loss rates. We have found striking evidence of a dichotomy between stars with ``normal'' 
field strengths and derived torques and those with low field strengths and small derived 
torques. Prior to \lamser, these categories corresponded well to expectations for a 
disrupted magnetic braking model. At first glance, \lamser appears to be an exception: it 
is in the ``normal'' field and torque state, but has a Rossby number beyond that 
predicted by a simple cutoff model. This intriguing result has a number of potential 
causes---we will begin with those consistent with the disrupted braking hypothesis.

The simplest explanation is a mechanical error in the derived stellar properties. For 
example, if the overturn timescale were longer or if we adopted a different age, \lamser 
might line up with expectations. Although this is certainly possible, we consider it 
unlikely based on our error model. A second variant would be that \lamser has experienced 
an unusual angular momentum history---for example, either a stellar merger or engulfment 
of a large planet. These events are actually not unusual for low mass stars, and they 
would reset the rotation and activity ``clock'' such that more rapid rotation and higher 
activity could be expected. \cite{Andronov2006} estimated that of order 4\% of stars in 
this mass and age range are actually merger products. However, field merger products tend 
to be Li-poor \citep{Ryan2002}, while \lamser is Li-rich \citep{XingXing2012}. A 
stellar-mass merger is thus disfavored, but late engulfment of a giant planet could 
account for the Li abundance and induce significant spin-up, particularly as stars age 
and begin to leave the main sequence. Exact rates are uncertain, but this cannot be ruled 
out at the few percent level \citep{Ahuir2021}. If apparent counter-examples like \lamser 
are rare, these explanations---errors in stellar measurements or an unusual 
history---would become more plausible.

A different family of solutions focuses instead on the nature of the threshold 
transition. Although a Rossby formulation is widely used in activity studies, it may be 
inadequate to capture the full picture. We are using evolutionary models of evolved 
stars, while traditional Rossby studies are confined to unevolved near-MS stars. As a 
result, traditional ``Rossby'' scaling can also be viewed as expressing a dependence of 
stellar activity on, say, effective temperature. The alignment with theoretical overturn 
timescales could be a happy coincidence. It may therefore be helpful to revisit the 
question of whether Rossby number really does serve as a valid dynamo diagnostic in the 
evolved and low activity domain. A variant of this hypothesis would be a duty cycle 
argument---in such a model, the transition from low to high state is not abrupt, but 
instead gradual. An example in the history of the Sun would be the existence of a Maunder 
minimum phase. In the transition domain between the active and inactive branches, stars 
would cycle between active and inactive phases over an extended period of time 
\citep{Vashishth2023}. Such a pattern could be revealed with a larger sample of stars, 
sufficient to infer statistically significant samples for hypothesis testing.

The evolutionary status of \lamser is ambiguous, at least in part, due to a weak 
constraint on the stellar age from asteroseismology. The age can be constrained from the 
frequency difference between radial ($l=0$) oscillation modes that sample the composition 
in the stellar core and neighboring quadrupole ($l=2$) modes that do not pass through 
this region. The non-detection of $l=2$ modes in \lamser is unusual for stars in this 
temperature range \citep{lund17b}, and prevents a determination of the small frequency 
separation ($\delta\nu_{02}$) that would otherwise provide a stronger constraint on the 
stellar age. Unfortunately, additional TESS observations of \lamser will not be available 
until 2026 at the earliest, because the position of Sector 78 was shifted northward to 
avoid scattered light from the Earth and Moon. However, ground-based radial velocity 
observations are less impacted by the background noise from stellar granulation 
\citep{GarciaBallot2019}, yielding a higher S/N than photometry and improving the 
potential to detect low-amplitude oscillation modes. Future observations of \lamser by 
the Stellar Observations Network Group \citep[SONG;][]{Grundahl2008} may provide a 
measurement of $\delta\nu_{02}$ that could substantially improve the age precision and 
help resolve this ambiguity.

\vspace*{12pt}
This paper includes data collected with the TESS mission, obtained from the Mikulski Archive for Space Telescopes (MAST) at the Space Telescope Science Institute (STScI). The specific observations analyzed can be accessed via \dataset[doi:10.17909/ar5x-2g05]{https://doi.org/10.17909/ar5x-2g05}. Funding for the TESS mission is provided by the NASA Explorer Program. STScI is operated by the Association of Universities for Research in Astronomy, Inc., under NASA contract NAS 5–26555.
T.S.M.\ acknowledges support from Chandra award GO0-21005X, NASA grant 80NSSC22K0475, NSF grant AST-2205919, and the Vanderbilt Initiative in Data-intensive Astrophysics (VIDA). Computational time at the Texas Advanced Computing Center was provided through XSEDE allocation TG-AST090107. 
D.B.\ gratefully acknowledges support from NASA (NNX16AB76G, 80NSSC22K0622) and the Whitaker Endowed Fund at Florida Gulf Coast University. 
D.H.\ acknowledges support from the Alfred P. Sloan Foundation, NASA (80NSSC22K0303, 80NSSC23K0434, 80NSSC23K0435), and the Australian Research Council (FT200100871).
J.v.S acknowledges support from NSF grant AST-2205919.
S.B.\ acknowledges NSF grant AST-2205026. 
O.K.\ acknowledges support by the Swedish Research Council (grant agreement no. 2019-03548), the Swedish National Space Agency, and the Royal Swedish Academy of Sciences. 
S.H.S.\ is grateful for support from award HST-GO-15991.002-A. 
V.S.\ acknowledges support from the European Space Agency (ESA) as an ESA Research Fellow. 
T.R.B.\ acknowledges support from the Australian Research Council (Laureate Fellowship FL220100117). 
S.N.B.\ acknowledges support from PLATO ASI-INAF agreement n.~2015-019-R.1-2018. 
A.J.F.\ acknowledges support from the European Research Council (ERC) under the European Union’s Horizon 2020 research and innovation programme (grant agreement No 810218 WHOLESUN).
R.A.G.\ acknowledges support from the PLATO and GOLF Centre National D'{\'{E}}tudes Spatiales (CNES) grant. 
Funding for the Stellar Astrophysics Centre is provided by The Danish National Research Foundation (Grant agreement No.~DNRF106). 
M.B.N.\ acknowledges support from the UK Space Agency. 
A.S.\ acknowledges support from the European Research Council Consolidator Grant funding scheme (project ASTEROCHRONOMETRY, G.A.\ n.\ 772293, http://www.asterochronometry.eu). 
C.A.C.\ acknowledges that this research was carried out at the Jet Propulsion Laboratory, California Institute of Technology, under a contract with NASA (80NM0018D0004). 
D.G.R.\ acknowledges support from the Spanish Ministry of Science and Innovation (MICINN) grant no. PID2019-107187GB-I00. 
K.G.S.\ thanks the German Federal Ministry (BMBF) for the year-long support for the construction of PEPSI through their Verbundforschung grants 05AL2BA1/3 and 05A08BAC as well as the State of Brandenburg for the continuing support of the LBT (see https://pepsi.aip.de/). LBT Corporation partners are the University of Arizona on behalf of the Arizona university system; Istituto Nazionale di Astrofisica, Italy; LBT Beteiligungsgesellschaft, Germany, representing the Max-Planck Society, the Leibniz-Institute for Astrophysics Potsdam (AIP), and Heidelberg University; the Ohio State University; and the Research Corporation, on behalf of the University of Notre Dame, University of Minnesota and University of Virginia.
S.V.J.\ acknowledges the support of the DFG priority program SPP 1992 ``Exploring the Diversity of Extrasolar Planets'' (JE 701/5-1).
A.A.V.\ acknowledges funding from the European Research Council (ERC) under the European Union's Horizon 2020 research and innovation programme (grant agreement No 817540, ASTROFLOW).

\appendix\vspace*{-24pt}
\section{Asteroseismic Non-detections from TESS}\label{appA} 

Using the same procedure outlined in Section \ref{sec2.1}, we produced light curves for 
targets that both had 20~s cadence data and fell along the evolutionary sequences for our 
spectropolarimetric targets (Table~\ref{tab6}). For these eight targets, we were able to 
improve on the quality of the light curve from the SPOC product. The light curves were 
analyzed for oscillations using pySYD \citep{chontos22, Huber2009}, yielding a null 
detection in all cases.

\begin{figure*}
\centering\includegraphics[width=\textwidth,trim=10 15 10 0,clip]{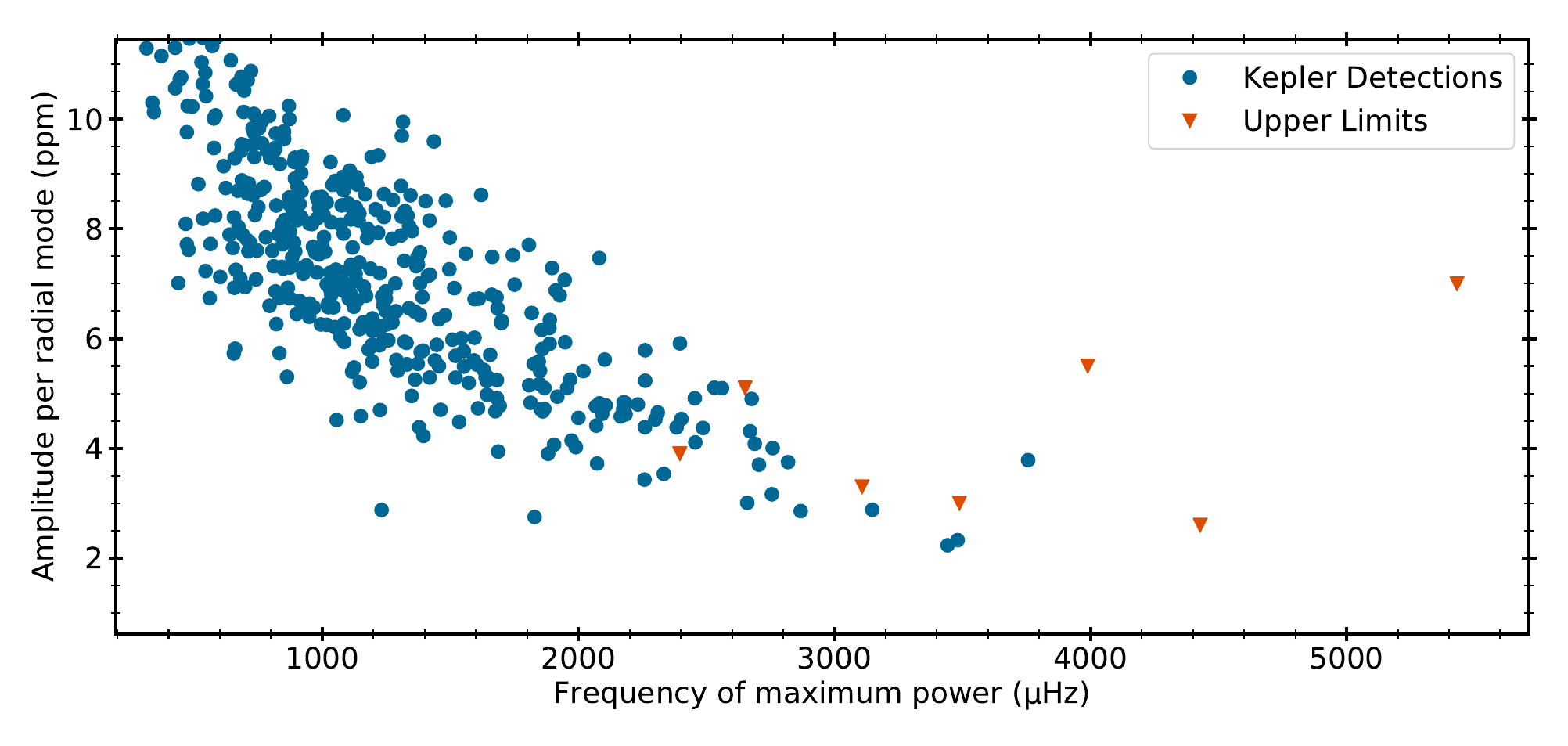}
\caption{Amplitude per radial mode for stars observed by Kepler \citep[blue 
circles;][]{huber11} compared to the upper amplitude limits derived from TESS for stars 
with 20~s cadence data that fall along evolutionary sequences for our spectropolarimetric 
targets (orange triangles). Note that the upper limit for $\epsilon$ Eri is off-scale at 
30.5, far above the markers in the legend.\label{fig:limits}}
\end{figure*}

To derive upper limits, we evaluated the fitted background model from pySYD at the 
predicted \numax\ value for each target and required a height-to-background ratio of 1.1 
\citep{mosser11c}, which is typically sufficient for a detection of oscillations. The 
corresponding amplitude limits are listed in Table~\ref{tab6} and compared to Kepler 
detections from \citet{huber11} in Figure~\ref{fig:limits}. To account for differences in 
the TESS and Kepler bandpass, we reduced the amplitude limits by a factor 0.8 
\citep{campante16}. The derived limits are consistent with null-detections for all stars. 
For the two stars with the lowest predicted \numax\ ($\iota$ Hor and 88~Leo) the limits 
are below some Kepler detections, implying that their amplitudes may be suppressed by 
stellar magnetic activity \citep{Garcia2010, chaplin11c, Mathur2019}.

\begin{deluxetable*}{cccccccc}
  \tablecaption{Upper limits on the detection of solar-like oscillations with TESS\label{tab6}}
  \tablehead{\colhead{Target} & \colhead{HD} & \colhead{Sector(s)} & \colhead{$T_{\rm eff}$} & \colhead{$\log g$} & \colhead{[M/H]} & \colhead{$\nu_{\rm max}$ ($\mu$Hz)} & \colhead{$A_{\rm lim}$ (ppm)}}
  \startdata
  $\iota$ Hor & 17051 & 30 & 6097 & 4.34 & +0.09 & 2396 & 3.9\\
  $\kappa^1$ Cet & 20630 & 31 & 5742 & 4.49 & +0.10 & 3488 & 3.0\\
  $\epsilon$ Eri & 22049 & 31 & 5146 & 4.57 & \phantom{+}0.00 & 4430 & 30.5\\ 
  40~Eri & 26965 & 32 & 5151 & 4.57 & $-$0.08 & 4428 & 2.6\\
  HD~76151 & 76151 & 34 & 5790 & 4.55 & +0.07 & 3989 & 5.5\\
  88~Leo & 100180 & 45, 49 & 5989 & 4.38 & $-$0.02 & 2651 & 5.1\\
  61~UMa & 101501 & 49 & 5488 & 4.43 & $-$0.03 & 3108 & 3.3\\
  HD~103095 & 103095 & 49 & 4950 & 4.65 & $-$1.16 & 5431 & 7.0\\
  \enddata
\end{deluxetable*}



\begin{thebibliography}{}
\expandafter\ifx\csname natexlab\endcsname\relax\def\natexlab#1{#1}\fi
\providecommand{\url}[1]{\href{#1}{#1}}
\providecommand{\dodoi}[1]{doi:~\href{http://doi.org/#1}{\nolinkurl{#1}}}
\providecommand{\doeprint}[1]{\href{http://ascl.net/#1}{\nolinkurl{http://ascl.net/#1}}}
\providecommand{\doarXiv}[1]{\href{https://arxiv.org/abs/#1}{\nolinkurl{https://arxiv.org/abs/#1}}}

\bibitem[{{Adelberger} {et~al.}(1998){Adelberger}, {Austin}, {Bahcall},
  {Balantekin}, {Bogaert}, {Brown}, {Buchmann}, {Cecil}, {Champagne}, {de
  Braeckeleer}, {Duba}, {Elliott}, {Freedman}, {Gai}, {Goldring}, {Gould},
  {Gruzinov}, {Haxton}, {Heeger}, {Henley}, {Johnson}, {Kamionkowski},
  {Kavanagh}, {Koonin}, {Kubodera}, {Langanke}, {Motobayashi}, {Pandharipande},
  {Parker}, {Robertson}, {Rolfs}, {Sawyer}, {Shaviv}, {Shoppa}, {Snover},
  {Swanson}, {Tribble}, {Turck-Chi{\`e}ze}, \&
  {Wilkerson}}]{1998RvMP...70.1265A}
{Adelberger}, E.~G., {Austin}, S.~M., {Bahcall}, J.~N., {et~al.} 1998, Reviews
  of Modern Physics, 70, 1265, \dodoi{10.1103/RevModPhys.70.1265}

\bibitem[{{Aguirre B{\o}rsen-Koch} {et~al.}(2022){Aguirre B{\o}rsen-Koch},
  {R{\o}rsted}, {Justesen}, {Stokholm}, {Verma}, {Winther}, {Knudstrup},
  {Nielsen}, {Sahlholdt}, {Larsen}, {Cassisi}, {Serenelli}, {Casagrande},
  {Christensen-Dalsgaard}, {Davies}, {Ferguson}, {Lund}, {Weiss}, \&
  {White}}]{Aguirre2022}
{Aguirre B{\o}rsen-Koch}, V., {R{\o}rsted}, J.~L., {Justesen}, A.~B., {et~al.}
  2022, \mnras, 509, 4344, \dodoi{10.1093/mnras/stab2911}

\bibitem[{{Ahuir} {et~al.}(2021){Ahuir}, {Strugarek}, {Brun}, \&
  {Mathis}}]{Ahuir2021}
{Ahuir}, J., {Strugarek}, A., {Brun}, A.~S., \& {Mathis}, S. 2021, \aap, 650,
  A126, \dodoi{10.1051/0004-6361/202040173}

\bibitem[{{Andronov} {et~al.}(2006){Andronov}, {Pinsonneault}, \&
  {Terndrup}}]{Andronov2006}
{Andronov}, N., {Pinsonneault}, M.~H., \& {Terndrup}, D.~M. 2006, \apj, 646,
  1160, \dodoi{10.1086/505127}

\bibitem[{{Antia} \& {Basu}(1994)}]{antia}
{Antia}, H.~M., \& {Basu}, S. 1994, \aaps, 107, 421

\bibitem[{{Appourchaux} {et~al.}(2012){Appourchaux}, {Benomar}, {Gruberbauer},
  {Chaplin}, {Garc{\'{\i}}a}, {Handberg}, {Verner}, {Antia}, {Campante},
  {Davies}, {Deheuvels}, {Hekker}, {Howe}, {Salabert}, {Bedding}, {White},
  {Houdek}, {Silva Aguirre}, {Elsworth}, {van Cleve}, {Clarke}, {Hall}, \&
  {Kjeldsen}}]{appourchaux12}
{Appourchaux}, T., {Benomar}, O., {Gruberbauer}, M., {et~al.} 2012, \aap, 537,
  A134, \dodoi{10.1051/0004-6361/201118496}

\bibitem[{{Auri{\`e}re}(2003)}]{Auriere2003}
{Auri{\`e}re}, M. 2003, in EAS Publications Series, Vol.~9, EAS Publications
  Series, ed. J.~{Arnaud} \& N.~{Meunier}, 105

\bibitem[{{Ayres} \& {Buzasi}(2022)}]{Ayres2022}
{Ayres}, T., \& {Buzasi}, D. 2022, \apjs, 263, 41,
  \dodoi{10.3847/1538-4365/ac8cfc}

\bibitem[{{Ayres}(2020)}]{Ayres2020}
{Ayres}, T.~R. 2020, \apjs, 250, 16, \dodoi{10.3847/1538-4365/aba3c6}

\bibitem[{{Baliunas} {et~al.}(1996){Baliunas}, {Sokoloff}, \&
  {Soon}}]{Baliunas1996}
{Baliunas}, S., {Sokoloff}, D., \& {Soon}, W. 1996, \apjl, 457, L99,
  \dodoi{10.1086/309891}

\bibitem[{{Baliunas} {et~al.}(1983){Baliunas}, {Hartmann}, {Noyes}, {Vaughan},
  {Preston}, {Frazer}, {Lanning}, {Middelkoop}, \& {Mihalas}}]{Baliunas1983}
{Baliunas}, S.~L., {Hartmann}, L., {Noyes}, R.~W., {et~al.} 1983, \apj, 275,
  752, \dodoi{10.1086/161572}

\bibitem[{{Baliunas} {et~al.}(1995){Baliunas}, {Donahue}, {Soon}, {Horne},
  {Frazer}, {Woodard-Eklund}, {Bradford}, {Rao}, {Wilson}, {Zhang}, {Bennett},
  {Briggs}, {Carroll}, {Duncan}, {Figueroa}, {Lanning}, {Misch}, {Mueller},
  {Noyes}, {Poppe}, {Porter}, {Robinson}, {Russell}, {Shelton}, {Soyumer},
  {Vaughan}, \& {Whitney}}]{Baliunas1995}
{Baliunas}, S.~L., {Donahue}, R.~A., {Soon}, W.~H., {et~al.} 1995, \apj, 438,
  269, \dodoi{10.1086/175072}

\bibitem[{{Ball} \& {Gizon}(2014)}]{ball_correction_2014}
{Ball}, W.~H., \& {Gizon}, L. 2014, \aap, 568, A123,
  \dodoi{10.1051/0004-6361/201424325}

\bibitem[{{Baum} {et~al.}(2022){Baum}, {Wright}, {Luhn}, \&
  {Isaacson}}]{Baum2022}
{Baum}, A.~C., {Wright}, J.~T., {Luhn}, J.~K., \& {Isaacson}, H. 2022, \aj,
  163, 183, \dodoi{10.3847/1538-3881/ac5683}

\bibitem[{{Bedding} {et~al.}(2007){Bedding}, {Kjeldsen}, {Arentoft}, {Bouchy},
  {Brandbyge}, {Brewer}, {Butler}, {Christensen-Dalsgaard}, {Dall}, {Frandsen},
  {Karoff}, {Kiss}, {Monteiro}, {Pijpers}, {Teixeira}, {Tinney}, {Baldry},
  {Carrier}, \& {O'Toole}}]{bedding07}
{Bedding}, T.~R., {Kjeldsen}, H., {Arentoft}, T., {et~al.} 2007, \apj, 663,
  1315, \dodoi{10.1086/518593}

\bibitem[{{Borucki} {et~al.}(2010){Borucki}, {Koch}, {Basri}, {Batalha},
  {Brown}, {Caldwell}, {Caldwell}, {Christensen-Dalsgaard}, {Cochran},
  {DeVore}, {Dunham}, {Dupree}, {Gautier}, {Geary}, {Gilliland}, {Gould},
  {Howell}, {Jenkins}, {Kondo}, {Latham}, {Marcy}, {Meibom}, {Kjeldsen},
  {Lissauer}, {Monet}, {Morrison}, {Sasselov}, {Tarter}, {Boss}, {Brownlee},
  {Owen}, {Buzasi}, {Charbonneau}, {Doyle}, {Fortney}, {Ford}, {Holman},
  {Seager}, {Steffen}, {Welsh}, {Rowe}, {Anderson}, {Buchhave}, {Ciardi},
  {Walkowicz}, {Sherry}, {Horch}, {Isaacson}, {Everett}, {Fischer}, {Torres},
  {Johnson}, {Endl}, {MacQueen}, {Bryson}, {Dotson}, {Haas}, {Kolodziejczak},
  {Van Cleve}, {Chandrasekaran}, {Twicken}, {Quintana}, {Clarke}, {Allen},
  {Li}, {Wu}, {Tenenbaum}, {Verner}, {Bruhweiler}, {Barnes}, \&
  {Prsa}}]{Borucki2010}
{Borucki}, W.~J., {Koch}, D., {Basri}, G., {et~al.} 2010, Science, 327, 977,
  \dodoi{10.1126/science.1185402}

\bibitem[{{Brage} {et~al.}(1996){Brage}, {Judge}, \& {Brekke}}]{Brage1996}
{Brage}, T., {Judge}, P.~G., \& {Brekke}, P. 1996, \apj, 464, 1030,
  \dodoi{10.1086/177390}

\bibitem[{{Breton} {et~al.}(2022){Breton}, {Garc{\'\i}a}, {Ballot}, {Delsanti},
  \& {Salabert}}]{breton22}
{Breton}, S.~N., {Garc{\'\i}a}, R.~A., {Ballot}, J., {Delsanti}, V., \&
  {Salabert}, D. 2022, \aap, 663, A118, \dodoi{10.1051/0004-6361/202243330}

\bibitem[{{Brewer} {et~al.}(2016){Brewer}, {Fischer}, {Valenti}, \&
  {Piskunov}}]{Brewer2016}
{Brewer}, J.~M., {Fischer}, D.~A., {Valenti}, J.~A., \& {Piskunov}, N. 2016,
  \apjs, 225, 32, \dodoi{10.3847/0067-0049/225/2/32}

\bibitem[{{Buzasi} {et~al.}(2015){Buzasi}, {Carboneau}, {Hessler}, {Lezcano},
  \& {Preston}}]{Buzasi2015}
{Buzasi}, D.~L., {Carboneau}, L., {Hessler}, C., {Lezcano}, A., \& {Preston},
  H. 2015, in IAU General Assembly, Vol.~29, 2256843

\bibitem[{{Campante} {et~al.}(2016){Campante}, {Schofield}, {Kuszlewicz},
  {Bouma}, {Chaplin}, {Huber}, {Christensen-Dalsgaard}, {Kjeldsen}, {Bossini},
  {North}, {Appourchaux}, {Latham}, {Pepper}, {Ricker}, {Stassun},
  {Vanderspek}, \& {Winn}}]{campante16}
{Campante}, T.~L., {Schofield}, M., {Kuszlewicz}, J.~S., {et~al.} 2016, \apj,
  830, 138, \dodoi{10.3847/0004-637X/830/2/138}

\bibitem[{{Carlos} {et~al.}(2016){Carlos}, {Nissen}, \&
  {Mel{\'e}ndez}}]{Carlos2016}
{Carlos}, M., {Nissen}, P.~E., \& {Mel{\'e}ndez}, J. 2016, \aap, 587, A100,
  \dodoi{10.1051/0004-6361/201527478}

\bibitem[{{Chaplin} {et~al.}(2011){Chaplin}, {Bedding}, {Bonanno}, {Broomhall},
  {Garc{\'{\i}}a}, {Hekker}, {Huber}, {Verner}, {Basu}, {Elsworth}, {Houdek},
  {Mathur}, {Mosser}, {New}, {Stevens}, {Appourchaux}, {Karoff}, {Metcalfe},
  {Molenda-{\.Z}akowicz}, {Monteiro}, {Thompson}, {Christensen-Dalsgaard},
  {Gilliland}, {Kawaler}, {Kjeldsen}, {Ballot}, {Benomar}, {Corsaro},
  {Campante}, {Gaulme}, {Hale}, {Handberg}, {Jarvis}, {R{\'e}gulo}, {Roxburgh},
  {Salabert}, {Stello}, {Mullally}, {Li}, \& {Wohler}}]{chaplin11c}
{Chaplin}, W.~J., {Bedding}, T.~R., {Bonanno}, A., {et~al.} 2011, \apjl, 732,
  L5, \dodoi{10.1088/2041-8205/732/1/L5}

\bibitem[{{Chontos} {et~al.}(2022){Chontos}, {Huber}, {Sayeed}, \&
  {Yamsiri}}]{chontos22}
{Chontos}, A., {Huber}, D., {Sayeed}, M., \& {Yamsiri}, P. 2022, The Journal of
  Open Source Software, 7, 3331, \dodoi{10.21105/joss.03331}

\bibitem[{{Christensen-Dalsgaard}(2008)}]{JCD08}
{Christensen-Dalsgaard}, J. 2008, \apss, 316, 13,
  \dodoi{10.1007/s10509-007-9675-5}

\bibitem[{{Corsaro} {et~al.}(2021){Corsaro}, {Bonanno}, {Mathur},
  {Garc{\'\i}a}, {Santos}, {Breton}, \& {Khalatyan}}]{Corsaro2021}
{Corsaro}, E., {Bonanno}, A., {Mathur}, S., {et~al.} 2021, \aap, 652, L2,
  \dodoi{10.1051/0004-6361/202141395}

\bibitem[{{Corsaro} \& {De Ridder}(2014)}]{corsaro14}
{Corsaro}, E., \& {De Ridder}, J. 2014, \aap, 571, A71,
  \dodoi{10.1051/0004-6361/201424181}

\bibitem[{{Corsaro} {et~al.}(2015){Corsaro}, {De Ridder}, \&
  {Garc{\'{\i}}a}}]{corsaro15}
{Corsaro}, E., {De Ridder}, J., \& {Garc{\'{\i}}a}, R.~A. 2015, \aap, 579, A83,
  \dodoi{10.1051/0004-6361/201525895}

\bibitem[{{Demarque} {et~al.}(2008){Demarque}, {Guenther}, {Li}, {Mazumdar}, \&
  {Straka}}]{demarque_yrec_2008}
{Demarque}, P., {Guenther}, D.~B., {Li}, L.~H., {Mazumdar}, A., \& {Straka},
  C.~W. 2008, \apss, 316, 31, \dodoi{10.1007/s10509-007-9698-y}

\bibitem[{{Donahue}(1993)}]{Donahue1993}
{Donahue}, R.~A. 1993, PhD thesis, New Mexico State University

\bibitem[{{Donahue} {et~al.}(1996){Donahue}, {Saar}, \&
  {Baliunas}}]{Donahue1996}
{Donahue}, R.~A., {Saar}, S.~H., \& {Baliunas}, S.~L. 1996, \apj, 466, 384,
  \dodoi{10.1086/177517}

\bibitem[{{Donati} {et~al.}(1997){Donati}, {Semel}, {Carter}, {Rees}, \&
  {Collier Cameron}}]{Donati1997}
{Donati}, J.-F., {Semel}, M., {Carter}, B.~D., {Rees}, D.~E., \& {Collier
  Cameron}, A. 1997, \mnras, 291, 658

\bibitem[{{Donati} {et~al.}(1992){Donati}, {Semel}, \& {Rees}}]{Donati1992}
{Donati}, J.~F., {Semel}, M., \& {Rees}, D.~E. 1992, \aap, 265, 669

\bibitem[{{Donati} {et~al.}(2006){Donati}, {Howarth}, {Jardine}, {Petit},
  {Catala}, {Landstreet}, {Bouret}, {Alecian}, {Barnes}, {Forveille},
  {Paletou}, \& {Manset}}]{Donati2006}
{Donati}, J.-F., {Howarth}, I.~D., {Jardine}, M.~M., {et~al.} 2006, MNRAS, 370,
  629, \dodoi{10.1111/j.1365-2966.2006.10558.x}

\bibitem[{{Egeland}(2017)}]{Egeland2017}
{Egeland}, R. 2017, PhD thesis, Montana State University, Bozeman, Montana, USA

\bibitem[{{Egeland} {et~al.}(2017){Egeland}, {Soon}, {Baliunas}, {Hall},
  {Pevtsov}, \& {Bertello}}]{Egeland2017b}
{Egeland}, R., {Soon}, W., {Baliunas}, S., {et~al.} 2017, \apj, 835,
  \dodoi{10.3847/1538-4357/835/1/25}

\bibitem[{{Ferguson} {et~al.}(2005){Ferguson}, {Alexander}, {Allard}, {Barman},
  {Bodnarik}, {Hauschildt}, {Heffner-Wong}, \& {Tamanai}}]{2005ApJ...623..585F}
{Ferguson}, J.~W., {Alexander}, D.~R., {Allard}, F., {et~al.} 2005, \apj, 623,
  585, \dodoi{10.1086/428642}

\bibitem[{{Finley} \& {Matt}(2018)}]{FinleyMatt2018}
{Finley}, A.~J., \& {Matt}, S.~P. 2018, \apj, 854, 78,
  \dodoi{10.3847/1538-4357/aaaab5}

\bibitem[{{Finley} {et~al.}(2018){Finley}, {Matt}, \& {See}}]{Finley2018}
{Finley}, A.~J., {Matt}, S.~P., \& {See}, V. 2018, \apj, 864, 125,
  \dodoi{10.3847/1538-4357/aad7b6}

\bibitem[{{Folsom} {et~al.}(2018{\natexlab{a}}){Folsom}, {Bouvier}, {Petit},
  {L{\`e}bre}, {Amard}, {Palacios}, {Morin}, {Donati}, \&
  {Vidotto}}]{Folsom2018a}
{Folsom}, C.~P., {Bouvier}, J., {Petit}, P., {et~al.} 2018{\natexlab{a}},
  \mnras, 474, 4956, \dodoi{10.1093/mnras/stx3021}

\bibitem[{{Folsom} {et~al.}(2018{\natexlab{b}}){Folsom}, {Fossati}, {Wood},
  {Sreejith}, {Cubillos}, {Vidotto}, {Alecian}, {Girish}, {Lichtenegger},
  {Murthy}, {Petit}, \& {Valyavin}}]{Folsom2018b}
{Folsom}, C.~P., {Fossati}, L., {Wood}, B.~E., {et~al.} 2018{\natexlab{b}},
  \mnras, 481, 5286, \dodoi{10.1093/mnras/sty2494}

\bibitem[{{Formicola} {et~al.}(2004){Formicola}, {Imbriani}, {Costantini},
  {Angulo}, {Bemmerer}, {Bonetti}, {Broggini}, {Corvisiero}, {Cruz},
  {Descouvemont}, {F{\"u}l{\"o}p}, {Gervino}, {Guglielmetti}, {Gustavino},
  {Gy{\"u}rky}, {Jesus}, {Junker}, {Lemut}, {Menegazzo}, {Prati}, {Roca},
  {Rolfs}, {Romano}, {Rossi Alvarez}, {Sch{\"u}mann}, {Somorjai}, {Straniero},
  {Strieder}, {Terrasi}, {Trautvetter}, {Vomiero}, \&
  {Zavatarelli}}]{2004PhLB..591...61F}
{Formicola}, A., {Imbriani}, G., {Costantini}, H., {et~al.} 2004, Physics
  Letters B, 591, 61, \dodoi{10.1016/j.physletb.2004.03.092}

\bibitem[{{Fruscione} {et~al.}(2006){Fruscione}, {McDowell}, {Allen},
  {Brickhouse}, {Burke}, {Davis}, {Durham}, {Elvis}, {Galle}, {Harris},
  {Huenemoerder}, {Houck}, {Ishibashi}, {Karovska}, {Nicastro}, {Noble},
  {Nowak}, {Primini}, {Siemiginowska}, {Smith}, \& {Wise}}]{Fruscione2006}
{Fruscione}, A., {McDowell}, J.~C., {Allen}, G.~E., {et~al.} 2006, in Society
  of Photo-Optical Instrumentation Engineers (SPIE) Conference Series, Vol.
  6270, Society of Photo-Optical Instrumentation Engineers (SPIE) Conference
  Series, ed. D.~R. {Silva} \& R.~E. {Doxsey}, 62701V

\bibitem[{{Gai} {et~al.}(2011){Gai}, {Basu}, {Chaplin}, \& {Elsworth}}]{YB}
{Gai}, N., {Basu}, S., {Chaplin}, W.~J., \& {Elsworth}, Y. 2011, \apj, 730, 63,
  \dodoi{10.1088/0004-637X/730/2/63}

\bibitem[{{Garc{\'\i}a} \& {Ballot}(2019)}]{GarciaBallot2019}
{Garc{\'\i}a}, R.~A., \& {Ballot}, J. 2019, Living Reviews in Solar Physics,
  16, 4, \dodoi{10.1007/s41116-019-0020-1}

\bibitem[{{Garc{\'\i}a} {et~al.}(2010){Garc{\'\i}a}, {Mathur}, {Salabert},
  {Ballot}, {R{\'e}gulo}, {Metcalfe}, \& {Baglin}}]{Garcia2010}
{Garc{\'\i}a}, R.~A., {Mathur}, S., {Salabert}, D., {et~al.} 2010, Science,
  329, 1032, \dodoi{10.1126/science.1191064}

\bibitem[{{Garc{\'{\i}}a} {et~al.}(2009){Garc{\'{\i}}a}, {R{\'e}gulo},
  {Samadi}, {Ballot}, {Barban}, {Benomar}, {Chaplin}, {Gaulme}, {Appourchaux},
  {Mathur}, {Mosser}, {Toutain}, {Verner}, {Auvergne}, {Baglin}, {Baudin},
  {Boumier}, {Bruntt}, {Catala}, {Deheuvels}, {Elsworth}, {Jim{\'e}nez-Reyes},
  {Michel}, {P{\'e}rez Hern{\'a}ndez}, {Roxburgh}, \& {Salabert}}]{garcia09}
{Garc{\'{\i}}a}, R.~A., {R{\'e}gulo}, C., {Samadi}, R., {et~al.} 2009, \aap,
  506, 41, \dodoi{10.1051/0004-6361/200911910}

\bibitem[{{Garc{\'{\i}}a} {et~al.}(2011){Garc{\'{\i}}a}, {Hekker}, {Stello},
  {Guti{\'e}rrez-Soto}, {Handberg}, {Huber}, {Karoff}, {Uytterhoeven},
  {Appourchaux}, {Chaplin}, {Elsworth}, {Mathur}, {Ballot},
  {Christensen-Dalsgaard}, {Gilliland}, {Houdek}, {Jenkins}, {Kjeldsen},
  {McCauliff}, {Metcalfe}, {Middour}, {Molenda-Zakowicz}, {Monteiro}, {Smith},
  \& {Thompson}}]{garcia11}
{Garc{\'{\i}}a}, R.~A., {Hekker}, S., {Stello}, D., {et~al.} 2011, \mnras, 414,
  L6, \dodoi{10.1111/j.1745-3933.2011.01042.x}

\bibitem[{{Gregory} \& {Loredo}(1992)}]{GregoryLoredo1992}
{Gregory}, P.~C., \& {Loredo}, T.~J. 1992, \apj, 398, 146,
  \dodoi{10.1086/171844}

\bibitem[{{Grundahl} {et~al.}(2008){Grundahl},
  {Christensen{\textendash}Dalsgaard}, {Arentoft}, {Frandsen}, {Kjeldsen},
  {J{\o}rgensen}, \& {Kjaergaard}}]{Grundahl2008}
{Grundahl}, F., {Christensen{\textendash}Dalsgaard}, J., {Arentoft}, T.,
  {et~al.} 2008, Communications in Asteroseismology, 157, 273

\bibitem[{{Hall} {et~al.}(2007){Hall}, {Lockwood}, \& {Skiff}}]{Hall2007}
{Hall}, J.~C., {Lockwood}, G.~W., \& {Skiff}, B.~A. 2007, \aj, 133, 862,
  \dodoi{10.1086/510356}

\bibitem[{{Hall} {et~al.}(2021){Hall}, {Davies}, {van Saders}, {Nielsen},
  {Lund}, {Chaplin}, {Garc{\'\i}a}, {Amard}, {Breimann}, {Khan}, {See}, \&
  {Tayar}}]{Hall2021}
{Hall}, O.~J., {Davies}, G.~R., {van Saders}, J., {et~al.} 2021, Nature
  Astronomy, 5, 707, \dodoi{10.1038/s41550-021-01335-x}

\bibitem[{{Handberg} \& {Campante}(2011)}]{handberg11}
{Handberg}, R., \& {Campante}, T.~L. 2011, \aap, 527, A56,
  \dodoi{10.1051/0004-6361/201015451}

\bibitem[{{Horne} \& {Baliunas}(1986)}]{Horne1986}
{Horne}, J.~H., \& {Baliunas}, S.~L. 1986, \apj, 302, 757,
  \dodoi{10.1086/164037}

\bibitem[{{Huber} {et~al.}(2009){Huber}, {Stello}, {Bedding}, {Chaplin},
  {Arentoft}, {Quirion}, \& {Kjeldsen}}]{Huber2009}
{Huber}, D., {Stello}, D., {Bedding}, T.~R., {et~al.} 2009, Communications in
  Asteroseismology, 160, 74, \dodoi{10.48550/arXiv.0910.2764}

\bibitem[{{Huber} {et~al.}(2011){Huber}, {Bedding}, {Arentoft}, {Gruberbauer},
  {Guenther}, {Houdek}, {Kallinger}, {Kjeldsen}, {Matthews}, {Stello}, \&
  {Weiss}}]{huber11}
{Huber}, D., {Bedding}, T.~R., {Arentoft}, T., {et~al.} 2011, \apj, 731, 94,
  \dodoi{10.1088/0004-637X/731/2/94}

\bibitem[{{Huber} {et~al.}(2022){Huber}, {White}, {Metcalfe}, {Chontos},
  {Fausnaugh}, {Ho}, {Van Eylen}, {Ball}, {Basu}, {Bedding}, {Benomar},
  {Bossini}, {Breton}, {Buzasi}, {Campante}, {Chaplin},
  {Christensen-Dalsgaard}, {Cunha}, {Deal}, {Garc{\'\i}a}, {Garc{\'\i}a
  Mu{\~n}oz}, {Gehan}, {Gonz{\'a}lez-Cuesta}, {Jiang}, {Kayhan}, {Kjeldsen},
  {Lundkvist}, {Mathis}, {Mathur}, {Monteiro}, {Nsamba}, {Ong},
  {Pak{\v{s}}tien{\.{e}}}, {Serenelli}, {Silva Aguirre}, {Stassun}, {Stello},
  {Norgaard Stilling}, {Lykke Winther}, {Wu}, {Barclay}, {Daylan},
  {G{\"u}nther}, {Hermes}, {Jenkins}, {Latham}, {Levine}, {Ricker}, {Seager},
  {Shporer}, {Twicken}, {Vanderspek}, \& {Winn}}]{huber22}
{Huber}, D., {White}, T.~R., {Metcalfe}, T.~S., {et~al.} 2022, \aj, 163, 79,
  \dodoi{10.3847/1538-3881/ac3000}

\bibitem[{{Iglesias} \& {Rogers}(1996)}]{1996ApJ...464..943I}
{Iglesias}, C.~A., \& {Rogers}, F.~J. 1996, \apj, 464, 943,
  \dodoi{10.1086/177381}

\bibitem[{{Isaacson} \& {Fischer}(2010)}]{Isaacson2010}
{Isaacson}, H., \& {Fischer}, D. 2010, \apj, 725, 875,
  \dodoi{10.1088/0004-637X/725/1/875}

\bibitem[{{Joyce} \& {Chaboyer}(2018)}]{Joyce2018}
{Joyce}, M., \& {Chaboyer}, B. 2018, \apj, 864, 99,
  \dodoi{10.3847/1538-4357/aad464}

\bibitem[{{Judge}(2020)}]{Judge2020}
{Judge}, P.~G. 2020, \mnras, 491, 576, \dodoi{10.1093/mnras/stz3063}

\bibitem[{{Keenan} {et~al.}(1988){Keenan}, {Dufton}, {Aggarwal}, \&
  {Kingston}}]{Keenan1988}
{Keenan}, F.~P., {Dufton}, P.~L., {Aggarwal}, K.~M., \& {Kingston}, A.~E. 1988,
  \apj, 324, 1068, \dodoi{10.1086/165963}

\bibitem[{{Keenan} {et~al.}(1990){Keenan}, {Dufton}, \&
  {Kingston}}]{Keenan1990}
{Keenan}, F.~P., {Dufton}, P.~L., \& {Kingston}, A.~E. 1990, \apj, 353, 636,
  \dodoi{10.1086/168653}

\bibitem[{{Keenan} {et~al.}(1987){Keenan}, {Kingston}, \&
  {Dufton}}]{Keenan1987}
{Keenan}, F.~P., {Kingston}, A.~E., \& {Dufton}, P.~L. 1987, \mnras, 225, 859,
  \dodoi{10.1093/mnras/225.4.859}

\bibitem[{{Kjeldsen} {et~al.}(2005){Kjeldsen}, {Bedding}, {Butler},
  {Christensen-Dalsgaard}, {Kiss}, {McCarthy}, {Marcy}, {Tinney}, \&
  {Wright}}]{kjeldsen05}
{Kjeldsen}, H., {Bedding}, T.~R., {Butler}, R.~P., {et~al.} 2005, \apj, 635,
  1281, \dodoi{10.1086/497530}

\bibitem[{{Kochukhov}(2016)}]{Kochukhov2016}
{Kochukhov}, O. 2016, in Lecture Notes in Physics, Berlin Springer Verlag, ed.
  J.-P. {Rozelot} \& C.~{Neiner}, Vol. 914 (Springer, Berlin), 177

\bibitem[{{Kochukhov} {et~al.}(2014){Kochukhov}, {L{\"u}ftinger}, {Neiner},
  {Alecian}, \& {MiMeS Collaboration}}]{Kochukhov2014}
{Kochukhov}, O., {L{\"u}ftinger}, T., {Neiner}, C., {Alecian}, E., \& {MiMeS
  Collaboration}. 2014, \aap, 565, A83, \dodoi{10.1051/0004-6361/201423472}

\bibitem[{{Kochukhov} {et~al.}(2010){Kochukhov}, {Makaganiuk}, \&
  {Piskunov}}]{Kochukhov2010}
{Kochukhov}, O., {Makaganiuk}, V., \& {Piskunov}, N. 2010, \aap, 524, A5,
  \dodoi{10.1051/0004-6361/201015429}

\bibitem[{{Landi Degl'Innocenti}(1992)}]{deglinnocenti1992}
{Landi Degl'Innocenti}, E. 1992, {Magnetic field measurements.} (Cambridge
  University Press, Cambridge), 71

\bibitem[{{Lehtinen} {et~al.}(2022){Lehtinen}, {K{\"a}pyl{\"a}}, {Hackman},
  {Kochukhov}, {Willamo}, {Marsden}, {Jeffers}, {Henry}, \&
  {Jetsu}}]{Lehtinen2022}
{Lehtinen}, J.~J., {K{\"a}pyl{\"a}}, M.~J., {Hackman}, T., {et~al.} 2022, \aap,
  660, A141, \dodoi{10.1051/0004-6361/201936780}

\bibitem[{{Lenz} \& {Breger}(2005)}]{lenz05}
{Lenz}, P., \& {Breger}, M. 2005, Communications in Asteroseismology, 146, 53,
  \dodoi{10.1553/cia146s53}

\bibitem[{{Li} {et~al.}(2020){Li}, {Bedding}, {Li}, {Bi}, {Stello}, {Zhou}, \&
  {White}}]{liyg20}
{Li}, Y., {Bedding}, T.~R., {Li}, T., {et~al.} 2020, \mnras, 495, 2363,
  \dodoi{10.1093/mnras/staa1335}

\bibitem[{{Li} {et~al.}(2023){Li}, {Bedding}, {Stello}, {Huber}, {Hon},
  {Joyce}, {Li}, {Perkins}, {White}, {Zinn}, {Howard}, {Isaacson}, {Hey}, \&
  {Kjeldsen}}]{liyg23}
{Li}, Y., {Bedding}, T.~R., {Stello}, D., {et~al.} 2023, \mnras, 523, 916,
  \dodoi{10.1093/mnras/stad1445}

\bibitem[{{Lomb}(1976)}]{Lomb1976}
{Lomb}, N.~R. 1976, \apss, 39, 447, \dodoi{10.1007/BF00648343}

\bibitem[{{L{\'o}pez Ariste} {et~al.}(2022){L{\'o}pez Ariste}, {Georgiev},
  {Mathias}, {L{\`e}bre}, {Wavasseur}, {Josselin}, {Konstantinova-Antova}, \&
  {Roudier}}]{LopezAriste2022}
{L{\'o}pez Ariste}, A., {Georgiev}, S., {Mathias}, P., {et~al.} 2022, \aap,
  661, A91, \dodoi{10.1051/0004-6361/202142271}

\bibitem[{{Lund} {et~al.}(2017){Lund}, {Silva Aguirre}, {Davies}, {Chaplin},
  {Christensen-Dalsgaard}, {Houdek}, {White}, {Bedding}, {Ball}, {Huber},
  {Antia}, {Lebreton}, {Latham}, {Handberg}, {Verma}, {Basu}, {Casagrande},
  {Justesen}, {Kjeldsen}, \& {Mosumgaard}}]{lund17b}
{Lund}, M.~N., {Silva Aguirre}, V., {Davies}, G.~R., {et~al.} 2017, \apj, 835,
  172, \dodoi{10.3847/1538-4357/835/2/172}

\bibitem[{{Marsden} {et~al.}(2014){Marsden}, {Petit}, {Jeffers}, {Morin},
  {Fares}, {Reiners}, {do Nascimento}, {Auri{\`e}re}, {Bouvier}, {Carter},
  {Catala}, {Dintrans}, {Donati}, {Gastine}, {Jardine}, {Konstantinova-Antova},
  {Lanoux}, {Ligni{\`e}res}, {Morgenthaler}, {Ram{\`\i}rez-V{\`e}lez},
  {Th{\'e}ado}, {Van Grootel}, \& {BCool Collaboration}}]{Marsden2014}
{Marsden}, S.~C., {Petit}, P., {Jeffers}, S.~V., {et~al.} 2014, \mnras, 444,
  3517, \dodoi{10.1093/mnras/stu1663}

\bibitem[{{Mathis} \& {Neiner}(2015)}]{MathisNeiner2015}
{Mathis}, S., \& {Neiner}, C. 2015, in New Windows on Massive Stars, ed.
  G.~{Meynet}, C.~{Georgy}, J.~{Groh}, \& P.~{Stee}, Vol. 307, 420--425

\bibitem[{{Mathur} {et~al.}(2019){Mathur}, {Garc{\'\i}a}, {Bugnet}, {Santos},
  {Santiago}, \& {Beck}}]{Mathur2019}
{Mathur}, S., {Garc{\'\i}a}, R.~A., {Bugnet}, L., {et~al.} 2019, Frontiers in
  Astronomy and Space Sciences, 6, 46, \dodoi{10.3389/fspas.2019.00046}

\bibitem[{{McQuillan} {et~al.}(2014){McQuillan}, {Mazeh}, \&
  {Aigrain}}]{mcquillan14}
{McQuillan}, A., {Mazeh}, T., \& {Aigrain}, S. 2014, \apjs, 211, 24,
  \dodoi{10.1088/0067-0049/211/2/24}

\bibitem[{{Mermilliod}(2006)}]{Mermilliod:2006}
{Mermilliod}, J.~C. 2006, VizieR Online Data Catalog, II/168

\bibitem[{{Metcalfe} {et~al.}(2009){Metcalfe}, {Creevey}, \&
  {Christensen-Dalsgaard}}]{Metcalfe2009}
{Metcalfe}, T.~S., {Creevey}, O.~L., \& {Christensen-Dalsgaard}, J. 2009, \apj,
  699, 373, \dodoi{10.1088/0004-637X/699/1/373}

\bibitem[{{Metcalfe} {et~al.}(2019){Metcalfe}, {Kochukhov}, {Ilyin},
  {Strassmeier}, {Godoy-Rivera}, \& {Pinsonneault}}]{Metcalfe2019b}
{Metcalfe}, T.~S., {Kochukhov}, O., {Ilyin}, I.~V., {et~al.} 2019, \apjl, 887,
  L38, \dodoi{10.3847/2041-8213/ab5e48}

\bibitem[{{Metcalfe} {et~al.}(2020){Metcalfe}, {van Saders}, {Basu}, {Buzasi},
  {Chaplin}, {Egeland}, {Garcia}, {Gaulme}, {Huber}, {Reinhold}, {Schunker},
  {Stassun}, {Appourchaux}, {Ball}, {Bedding}, {Deheuvels},
  {Gonz{\'a}lez-Cuesta}, {Handberg}, {Jim{\'e}nez}, {Kjeldsen}, {Li}, {Lund},
  {Mathur}, {Mosser}, {Nielsen}, {Noll}, {{\c{C}}elik Orhan}, {{\"O}rtel},
  {Santos}, {Yildiz}, {Baliunas}, \& {Soon}}]{Metcalfe2020}
{Metcalfe}, T.~S., {van Saders}, J.~L., {Basu}, S., {et~al.} 2020, \apj, 900,
  154, \dodoi{10.3847/1538-4357/aba963}

\bibitem[{{Metcalfe} {et~al.}(2021){Metcalfe}, {van Saders}, {Basu}, {Buzasi},
  {Drake}, {Egeland}, {Huber}, {Saar}, {Stassun}, {Ball}, {Campante}, {Finley},
  {Kochukhov}, {Mathur}, {Reinhold}, {See}, {Baliunas}, \&
  {Soon}}]{Metcalfe2021}
---. 2021, \apj, 921, 122, \dodoi{10.3847/1538-4357/ac1f19}

\bibitem[{{Metcalfe} {et~al.}(2022){Metcalfe}, {Finley}, {Kochukhov}, {See},
  {Ayres}, {Stassun}, {van Saders}, {Clark}, {Godoy-Rivera}, {Ilyin},
  {Pinsonneault}, {Strassmeier}, \& {Petit}}]{Metcalfe2022}
{Metcalfe}, T.~S., {Finley}, A.~J., {Kochukhov}, O., {et~al.} 2022, \apjl, 933,
  L17, \dodoi{10.3847/2041-8213/ac794d}

\bibitem[{{Metcalfe} {et~al.}(2023){Metcalfe}, {Strassmeier}, {Ilyin}, {van
  Saders}, {Ayres}, {Finley}, {Kochukhov}, {Petit}, {See}, {Stassun},
  {Jeffers}, {Marsden}, {Morin}, \& {Vidotto}}]{Metcalfe2023}
{Metcalfe}, T.~S., {Strassmeier}, K.~G., {Ilyin}, I.~V., {et~al.} 2023, \apjl,
  948, L6, \dodoi{10.3847/2041-8213/acce38}

\bibitem[{{Morel}(2018)}]{Morel2018}
{Morel}, T. 2018, \aap, 615, A172, \dodoi{10.1051/0004-6361/201833125}

\bibitem[{{Mosser} {et~al.}(2012){Mosser}, {Elsworth}, {Hekker}, {Huber},
  {Kallinger}, {Mathur}, {Belkacem}, {Goupil}, {Samadi}, {Barban}, {Bedding},
  {Chaplin}, {Garc{\'{\i}}a}, {Stello}, {De Ridder}, {Middour}, {Morris}, \&
  {Quintana}}]{mosser11c}
{Mosser}, B., {Elsworth}, Y., {Hekker}, S., {et~al.} 2012, \aap, 537, A30,
  \dodoi{10.1051/0004-6361/201117352}

\bibitem[{Nason(2006)}]{Nason2006}
Nason, G. 2006, Stationary and non-stationary time series (United Kingdom:
  Geological Society of London), 129 -- 142

\bibitem[{{Nielsen} {et~al.}(2020){Nielsen}, {Ball}, {Standing}, {Triaud},
  {Buzasi}, {Carboneau}, {Stassun}, {Kane}, {Chaplin}, {Bellinger}, {Mosser},
  {Roxburgh}, {{\c{C}}elik Orhan}, {Y{\i}ld{\i}z}, {{\"O}rtel}, {Vrard},
  {Mazumdar}, {Ranadive}, {Deal}, {Davies}, {Campante}, {Garc{\'\i}a},
  {Mathur}, {Gonz{\'a}lez-Cuesta}, \& {Serenelli}}]{Nielsen2020}
{Nielsen}, M.~B., {Ball}, W.~H., {Standing}, M.~R., {et~al.} 2020, \aap, 641,
  A25, \dodoi{10.1051/0004-6361/202037461}

\bibitem[{{Paunzen}(2015)}]{Paunzen:2015}
{Paunzen}, E. 2015, \aap, 580, A23, \dodoi{10.1051/0004-6361/201526413}

\bibitem[{{Paxton} {et~al.}(2015){Paxton}, {Marchant}, {Schwab}, {Bauer},
  {Bildsten}, {Cantiello}, {Dessart}, {Farmer}, {Hu}, {Langer}, {Townsend},
  {Townsley}, \& {Timmes}}]{Paxton2015}
{Paxton}, B., {Marchant}, P., {Schwab}, J., {et~al.} 2015, \apjs, 220, 15,
  \dodoi{10.1088/0067-0049/220/1/15}

\bibitem[{{Petit} {et~al.}(2022){Petit}, {B{\"o}hm}, {Folsom}, {Ligni{\`e}res},
  \& {Cang}}]{Petit2022}
{Petit}, P., {B{\"o}hm}, T., {Folsom}, C.~P., {Ligni{\`e}res}, F., \& {Cang},
  T. 2022, \aap, 666, A20, \dodoi{10.1051/0004-6361/202143000}

\bibitem[{{Petit} {et~al.}(2002){Petit}, {Donati}, \& {Collier
  Cameron}}]{Petit2002}
{Petit}, P., {Donati}, J.-F., \& {Collier Cameron}, A. 2002, \mnras, 334, 374,
  \dodoi{10.1046/j.1365-8711.2002.05529.x}

\bibitem[{{Petit} {et~al.}(2008){Petit}, {Dintrans}, {Solanki}, {Donati},
  {Auri{\`e}re}, {Ligni{\`e}res}, {Morin}, {Paletou}, {Ramirez Velez},
  {Catala}, \& {Fares}}]{Petit2008}
{Petit}, P., {Dintrans}, B., {Solanki}, S.~K., {et~al.} 2008, \mnras, 388, 80,
  \dodoi{10.1111/j.1365-2966.2008.13411.x}

\bibitem[{{Petit} {et~al.}(2010){Petit}, {Ligni{\`e}res}, {Wade},
  {Auri{\`e}re}, {B{\"o}hm}, {Bagnulo}, {Dintrans}, {Fumel}, {Grunhut},
  {Lanoux}, {Morgenthaler}, \& {Van Grootel}}]{Petit2010}
{Petit}, P., {Ligni{\`e}res}, F., {Wade}, G.~A., {et~al.} 2010, \aap, 523, A41,
  \dodoi{10.1051/0004-6361/201015307}

\bibitem[{{Petit} {et~al.}(2021){Petit}, {Folsom}, {Donati}, {Yu}, {do
  Nascimento}, {Jeffers}, {Marsden}, {Morin}, \& {Vidotto}}]{Petit2021}
{Petit}, P., {Folsom}, C.~P., {Donati}, J.~F., {et~al.} 2021, \aap, 648, A55,
  \dodoi{10.1051/0004-6361/202040027}

\bibitem[{{Pr{\v{s}}a} {et~al.}(2019){Pr{\v{s}}a}, {Zhang}, \&
  {Wells}}]{Prsa2019}
{Pr{\v{s}}a}, A., {Zhang}, M., \& {Wells}, M. 2019, \pasp, 131, 068001,
  \dodoi{10.1088/1538-3873/ab0f41}

\bibitem[{{Rao} {et~al.}(2022){Rao}, {Del Zanna}, {Mason}, \&
  {Dufresne}}]{Rao2022}
{Rao}, Y.~K., {Del Zanna}, G., {Mason}, H.~E., \& {Dufresne}, R. 2022, \mnras,
  517, 1422, \dodoi{10.1093/mnras/stac2772}

\bibitem[{{Ricker} {et~al.}(2014){Ricker}, {Winn}, {Vanderspek}, {Latham},
  {Bakos}, {Bean}, {Berta-Thompson}, {Brown}, {Buchhave}, {Butler}, {Butler},
  {Chaplin}, {Charbonneau}, {Christensen-Dalsgaard}, {Clampin}, {Deming},
  {Doty}, {De Lee}, {Dressing}, {Dunham}, {Endl}, {Fressin}, {Ge}, {Henning},
  {Holman}, {Howard}, {Ida}, {Jenkins}, {Jernigan}, {Johnson}, {Kaltenegger},
  {Kawai}, {Kjeldsen}, {Laughlin}, {Levine}, {Lin}, {Lissauer}, {MacQueen},
  {Marcy}, {McCullough}, {Morton}, {Narita}, {Paegert}, {Palle}, {Pepe},
  {Pepper}, {Quirrenbach}, {Rinehart}, {Sasselov}, {Sato}, {Seager},
  {Sozzetti}, {Stassun}, {Sullivan}, {Szentgyorgyi}, {Torres}, {Udry}, \&
  {Villasenor}}]{Ricker2014}
{Ricker}, G.~R., {Winn}, J.~N., {Vanderspek}, R., {et~al.} 2014, in Society of
  Photo-Optical Instrumentation Engineers (SPIE) Conference Series, Vol. 9143,
  Proceedings of the SPIE, Volume 9143, id. 914320 15 pp. (2014)., 914320

\bibitem[{{Rogers} \& {Nayfonov}(2002)}]{nayfonov}
{Rogers}, F.~J., \& {Nayfonov}, A. 2002, \apj, 576, 1064,
  \dodoi{10.1086/341894}

\bibitem[{{Ros{\'e}n} {et~al.}(2016){Ros{\'e}n}, {Kochukhov}, {Hackman}, \&
  {Lehtinen}}]{Rosen2016}
{Ros{\'e}n}, L., {Kochukhov}, O., {Hackman}, T., \& {Lehtinen}, J. 2016, \aap,
  593, A35, \dodoi{10.1051/0004-6361/201628443}

\bibitem[{{Rosenthal} {et~al.}(2021){Rosenthal}, {Fulton}, {Hirsch},
  {Isaacson}, {Howard}, {Dedrick}, {Sherstyuk}, {Blunt}, {Petigura}, {Knutson},
  {Behmard}, {Chontos}, {Crepp}, {Crossfield}, {Dalba}, {Fischer}, {Henry},
  {Kane}, {Kosiarek}, {Marcy}, {Rubenzahl}, {Weiss}, \&
  {Wright}}]{Rosenthal2021}
{Rosenthal}, L.~J., {Fulton}, B.~J., {Hirsch}, L.~A., {et~al.} 2021, \apjs,
  255, 8, \dodoi{10.3847/1538-4365/abe23c}

\bibitem[{{Ryabchikova} {et~al.}(2015){Ryabchikova}, {Piskunov}, {Kurucz},
  {Stempels}, {Heiter}, {Pakhomov}, \& {Barklem}}]{Ryabchikova2015}
{Ryabchikova}, T., {Piskunov}, N., {Kurucz}, R.~L., {et~al.} 2015, \physscr,
  90, 054005, \dodoi{10.1088/0031-8949/90/5/054005}

\bibitem[{{Ryan} {et~al.}(2002){Ryan}, {Gregory}, {Kolb}, {Beers}, \&
  {Kajino}}]{Ryan2002}
{Ryan}, S.~G., {Gregory}, S.~G., {Kolb}, U., {Beers}, T.~C., \& {Kajino}, T.
  2002, \apj, 571, 501, \dodoi{10.1086/339939}

\bibitem[{{Scargle}(1982)}]{Scargle1982}
{Scargle}, J.~D. 1982, \apj, 263, 835, \dodoi{10.1086/160554}

\bibitem[{{Semel}(1989)}]{Semel1989}
{Semel}, M. 1989, A\&A, 225, 456

\bibitem[{{Semel} {et~al.}(1993){Semel}, {Donati}, \& {Rees}}]{Semel1993}
{Semel}, M., {Donati}, J.~F., \& {Rees}, D.~E. 1993, \aap, 278, 231

\bibitem[{{Soubiran} {et~al.}(2018){Soubiran}, {Jasniewicz}, {Chemin},
  {Zurbach}, {Brouillet}, {Panuzzo}, {Sartoretti}, {Katz}, {Le Campion},
  {Marchal}, {Hestroffer}, {Th{\'e}venin}, {Crifo}, {Udry}, {Cropper},
  {Seabroke}, {Viala}, {Benson}, {Blomme}, {Jean-Antoine}, {Huckle}, {Smith},
  {Baker}, {Damerdji}, {Dolding}, {Fr{\'e}mat}, {Gosset}, {Guerrier}, {Guy},
  {Haigron}, {Jan{\ss}en}, {Plum}, {Fabre}, {Lasne}, {Pailler}, {Panem},
  {Riclet}, {Royer}, {Tauran}, {Zwitter}, {Gueguen}, \& {Turon}}]{Soubiran2018}
{Soubiran}, C., {Jasniewicz}, G., {Chemin}, L., {et~al.} 2018, \aap, 616, A7,
  \dodoi{10.1051/0004-6361/201832795}

\bibitem[{{Stassun} {et~al.}(2017){Stassun}, {Collins}, \&
  {Gaudi}}]{Stassun:2017}
{Stassun}, K.~G., {Collins}, K.~A., \& {Gaudi}, B.~S. 2017, \aj, 153, 136,
  \dodoi{10.3847/1538-3881/aa5df3}

\bibitem[{{Stassun} {et~al.}(2018){Stassun}, {Corsaro}, {Pepper}, \&
  {Gaudi}}]{Stassun:2018}
{Stassun}, K.~G., {Corsaro}, E., {Pepper}, J.~A., \& {Gaudi}, B.~S. 2018, \aj,
  155, 22, \dodoi{10.3847/1538-3881/aa998a}

\bibitem[{{Stassun} \& {Torres}(2016)}]{Stassun:2016}
{Stassun}, K.~G., \& {Torres}, G. 2016, \aj, 152, 180,
  \dodoi{10.3847/0004-6256/152/6/180}

\bibitem[{{Stassun} \& {Torres}(2021)}]{StassunTorres:2021}
---. 2021, \apjl, 907, L33, \dodoi{10.3847/2041-8213/abdaad}

\bibitem[{{Strassmeier} {et~al.}(2015){Strassmeier}, {Ilyin}, {J{\"a}rvinen},
  {Weber}, {Woche}, {Barnes}, {Bauer}, {Beckert}, {Bittner}, {Bredthauer},
  {Carroll}, {Denker}, {Dionies}, {DiVarano}, {D{\"o}scher}, {Fechner},
  {Feuerstein}, {Granzer}, {Hahn}, {Harnisch}, {Hofmann}, {Lesser}, {Paschke},
  {Pankratow}, {Plank}, {Pl{\"u}schke}, {Popow}, \&
  {Sablowski}}]{Strassmeier2015}
{Strassmeier}, K.~G., {Ilyin}, I., {J{\"a}rvinen}, A., {et~al.} 2015,
  Astronomische Nachrichten, 336, 324, \dodoi{10.1002/asna.201512172}

\bibitem[{{Thoul} {et~al.}(1994){Thoul}, {Bahcall}, \&
  {Loeb}}]{thoul_element_1994}
{Thoul}, A.~A., {Bahcall}, J.~N., \& {Loeb}, A. 1994, \apj, 421, 828,
  \dodoi{10.1086/173695}

\bibitem[{{Torres} {et~al.}(2010){Torres}, {Andersen}, \&
  {Gim{\'e}nez}}]{Torres:2010}
{Torres}, G., {Andersen}, J., \& {Gim{\'e}nez}, A. 2010, \aapr, 18, 67,
  \dodoi{10.1007/s00159-009-0025-1}

\bibitem[{{van Saders} {et~al.}(2016){van Saders}, {Ceillier}, {Metcalfe},
  {Silva Aguirre}, {Pinsonneault}, {Garc{\'\i}a}, {Mathur}, \&
  {Davies}}]{vanSaders2016}
{van Saders}, J.~L., {Ceillier}, T., {Metcalfe}, T.~S., {et~al.} 2016, \nat,
  529, 181, \dodoi{10.1038/nature16168}

\bibitem[{{van Saders} {et~al.}(2019){van Saders}, {Pinsonneault}, \&
  {Barbieri}}]{vansaders19}
{van Saders}, J.~L., {Pinsonneault}, M.~H., \& {Barbieri}, M. 2019, \apj, 872,
  128, \dodoi{10.3847/1538-4357/aafafe}

\bibitem[{{Vanderburg} \& {Johnson}(2014)}]{Vanderburg2014}
{Vanderburg}, A., \& {Johnson}, J.~A. 2014, \pasp, 126, 948,
  \dodoi{10.1086/678764}

\bibitem[{{Vashishth} {et~al.}(2023){Vashishth}, {Karak}, \&
  {Kitchatinov}}]{Vashishth2023}
{Vashishth}, V., {Karak}, B.~B., \& {Kitchatinov}, L. 2023, \mnras, 522, 2601,
  \dodoi{10.1093/mnras/stad1105}

\bibitem[{{Vaughan} {et~al.}(1978){Vaughan}, {Preston}, \&
  {Wilson}}]{Vaughan1978}
{Vaughan}, A.~H., {Preston}, G.~W., \& {Wilson}, O.~C. 1978, \pasp, 90, 267,
  \dodoi{10.1086/130324}

\bibitem[{{Viani} \& {Basu}(2017)}]{viani_diffusion}
{Viani}, L., \& {Basu}, S. 2017, in European Physical Journal Web of
  Conferences, Vol. 160, European Physical Journal Web of Conferences, 05005

\bibitem[{{Weiss} {et~al.}(2008){Weiss}, {Moffat}, \& Kudelka}]{weiss08}
{Weiss}, W.~W., {Moffat}, A.~F.~J., \& Kudelka, O. 2008, Communications in
  Asteroseismology, 157, 271

\bibitem[{{White} \& {Livingston}(1981)}]{White1981}
{White}, O.~R., \& {Livingston}, W.~C. 1981, \apj, 249, 798,
  \dodoi{10.1086/159338}

\bibitem[{{White} {et~al.}(2011){White}, {Bedding}, {Stello},
  {Christensen-Dalsgaard}, {Huber}, \& {Kjeldsen}}]{white11}
{White}, T.~R., {Bedding}, T.~R., {Stello}, D., {et~al.} 2011, \apj, 743, 161,
  \dodoi{10.1088/0004-637X/743/2/161}

\bibitem[{{Wilson}(1978)}]{Wilson1978}
{Wilson}, O.~C. 1978, \apj, 226, 379, \dodoi{10.1086/156618}

\bibitem[{{Wood} {et~al.}(2018){Wood}, {Laming}, {Warren}, \&
  {Poppenhaeger}}]{Wood+2018}
{Wood}, B.~E., {Laming}, J.~M., {Warren}, H.~P., \& {Poppenhaeger}, K. 2018,
  \apj, 862, 66, \dodoi{10.3847/1538-4357/aaccf6}

\bibitem[{{Wood} {et~al.}(2021){Wood}, {M{\"u}ller}, {Redfield}, {Konow},
  {Vannier}, {Linsky}, {Youngblood}, {Vidotto}, {Jardine},
  {Alvarado-G{\'o}mez}, \& {Drake}}]{Wood2021}
{Wood}, B.~E., {M{\"u}ller}, H.-R., {Redfield}, S., {et~al.} 2021, \apj, 915,
  37, \dodoi{10.3847/1538-4357/abfda5}

\bibitem[{{Xing} \& {Xing}(2012)}]{XingXing2012}
{Xing}, L.~F., \& {Xing}, Q.~F. 2012, \aap, 537, A91,
  \dodoi{10.1051/0004-6361/201117133}

\end{thebibliography}
\end{document}